\documentclass[prb,reprint,10pt,aps]{revtex4-1}
\usepackage{amsmath,amssymb}
\usepackage{upgreek}
\usepackage{comment}
\usepackage{graphicx}

\usepackage{multirow}
\usepackage{here}

\begin{document}
	
	\title{Hot carrier dynamics and electron-optical phonon coupling in photoexcited graphene via time-resolved ultrabroadband terahertz spectroscopy }%

	\author{Sho Ikeda}%
	\author{Chiko Otani}%
	\author{Masatsugu Yamashita}%
	\email[e-mail: ]{m-yama@riken.jp}
	\affiliation{Terahertz Sensing and Imaging Team, RIKEN Center for Advanced Photonics, 519-1399 Aramaki-Aoba Aoba-ku, Sendai, Miyagi, 980-0845, Japan}
	\date{11, March 2021}%

	\begin{abstract}
	Electron-electron (e-e) interaction is known as a source of  logarithmic renormalizations for Dirac fermions in quantum field theory. The renormalization of electron--optical phonon coupling (EPC) by e-e interaction, which plays a pivotal role in hot carrier and phonon dynamics, has been discussed after the discovery of graphene. 
	We investigate the hot carrier dynamics and the EPC strength using time-resolved ultrabroadband terahertz (THz) spectroscopy combined with  numerical simulation based on the Boltzmann transport equation and comprehensive temperature model. The numerical simulation demonstrates that the extrinsic carrier scatterings by the Coulomb potential of the charged impurity and surface polar phonons are significantly suppressed by the carrier screening effect and have negligible contributions to the THz photoconductivity in heavily doped graphene on polyethylene terephthalate (PET) substrate. The large negative photoconductivity and the non-Drude behavior of THz conductivity spectra appear under high pump fluence and can be attributed to the temporal variation of the hot carrier distribution and scattering rate. The transient reflectivity well reflects the EPC strength and temporal evolution of the hot carrier and optical phonon dynamics. We successfully estimate the EPC matrix element of the $A_1^{\prime}$ optical phonon mode near the $\mathbf{K}$ point as $\left\langle D_{\mathbf{K}}^{2}\right\rangle_{\mathrm{F}} \approx 450\,(\mathrm{eV\AA^{-1}})^2$ from the fitting of THz conductivity spectra and temporal evolution of  transient THz reflectivity. The corresponding dimensionless EPC constant $\lambda_{\mathbf{K}} \approx$0.09 at Fermi energy $\varepsilon_{\mathrm{F}}=0.43\,\mathrm{eV}$ is slightly larger than the prediction of the renormalization group  approach including the dielectric screening effect of the PET substrate. This leads to the significant difference in hot carrier and phonon dynamics compared to those without the renormalization effect by the e-e interaction. This approach can provide a quantitative understanding of hot carrier and optical phonon dynamics, and support the development of future graphene optoelectronic devices.
	\end{abstract}
	
	\maketitle
	
	\section{Introduction}
	Hot carrier effects are regarded as insightful in studying many-body interactions in condensed matter, and play a crucial role in the operation of electronics and optoelectronic devices. For this reason, they have been investigated extensively in both metals and semiconductors\cite{DelFatti2000, Nozik2001}. The rise of  graphene had offered new opportunities for this research field because the carriers thereof are 2D massless Dirac fermions (MDFs) with a linear energy dispersion. This fact has promoted graphene as an attractive platform for hot carrier physics and various applications\cite{Xia2009, Tse2009, Berciaud2010, Xu2010, Gabor2011, Sun2012b, Wu2012, Liu2014, Song2015b, Tielrooij2015a, Stange2015, Kane2015,  McKitterick2015, Bonaccorso2015a, Li2019, Chen2019, Lin2019, Lin2019a, Kim2020, Kim2020a, Massicotte2021}.
	Electron or hole relaxation mainly involves non-radiative electron--electron (e-e) and electron--phonon scatterings, depending on the excitation energy.
	Electron--electron interaction is dominant at high energy, redistributes the electrical or optical power within the electron gas, and builds up a hot carrier population. 
	Electron--phonon interaction operates on  a longer time scale to equilibrate the electron and phonon temperatures, and to cool the hot carriers \cite{Pogna2021}. 
	
	Hot carrier effects play a significant role in the optoelectronic properties of photoexcited graphene, in which the photocarriers are excited at high energies. The subsequent relaxation drives the working efficiency of optoelectronic devices. In this respect, spectroscopic investigations such as pump probe spectroscopy\cite{Sun2008a} and angle-resolved photo-electron spectroscopy\cite{Gierz2013c, Johannsen2013b} of hot carriers complement transport studies. Optical pump terahertz (THz) probe spectroscopy (OPTP) is a powerful tool for investigating the hot carrier dynamics of graphene because it probes the intraband optical conductivity dominated not only by the hot carrier distribution, but also the carrier scattering process in contrast to optical pump optical probe spectroscopy. Extensive studies using OPTP \cite{George2008a, Strait2011a, Boubanga-Tombet2012a,Docherty2012e, Frenzel2013a, Jnawali2013b, Lin2013d,Tielrooij2013a, Frenzel2014d, Shi2014a, Jensen2014b, Kar2014a, Heyman2015a, Mihnev2016a} have revealed the unusual behaviors of graphene hot carriers, which undergo positive and negative changes in the intraband optical conductivity with non-Drude type frequency dependence. The negative change observed in heavily doped graphene is an indicative of enhanced carrier scattering and reduced Drude weight in quasi-equilibrium hot carrier state with a single chemical potential owing to ultrafast recombination of photoexcited carriers. However, most of these works were performed by THz probe with the relatively narrow band (1-3 THz) which was not sufficient for capturing the whole spectrum of non-Drude type conductivity  and their results have been interpreted using the framework of the phenomenological model \cite{Docherty2012e, Frenzel2014d, Shi2014a, Heyman2015a, Jnawali2013b}. Such a phenomenological analysis for the narrow band spectra is not sufficient to understand the hot carrier and phonon dynamics quantitatively and to derive the microscopic parameters. Theoretical studies have been conducted by incorporating the microscopic theory based on the density matrix formalism\cite{Mihnev2016a} or Boltzmann transport equation (BTE)\cite{Tomadin2018c, Yamashita2021}.
	
	The electron--optical phonon coupling (EPC) strength is a crucial factor that makes it difficult to understand the hot carrier and phonon relaxation process by numerical studies. The density functional theory (DFT) calculations demonstrated that only three strongly coupled optical phonon (SCOP) modes contribute significantly to the inelastic carrier scattering in graphene\cite{Piscanec2004a, Baroni2001}. 
	The first two relevant modes are associated with the G peak of the Raman spectrum and the highest optical branches at $\mathbf{\Gamma}$ (the $E_{2g}$ mode) with the energy of $\hbar \omega_{\mathrm{\bf{\Gamma}}}=196.0\,\mathrm{meV}$, which split into an upper longitudinal optical (LO) branch and a lower transverse optical (TO) branch near $\mathbf{\Gamma}$. Owing to their long wavelengths, these phonons scatter electrons within one valley. 
	Moreover, it is essential to take into account the highest optical branch of the zone boundary phonon $\hbar \omega_{\mathrm{\bf{K}}}=161\,\mathrm{meV}$ at the \textbf{K} point (the $A_1^{\prime}$ mode). This mode is responsible for intervalley processes and associated with the D and 2D peaks of the Raman spectrum. 
	In Refs. \cite{Piscanec2004a, Pisana2007a, Lazzeri2008, Calandra2007b}, the EPCs $\left\langle D_{\eta}^{2}\right\rangle_{\mathrm{F}}$ for dominant optical phonon modes $\eta$ ($\mathbf{\Gamma}_{\mathrm{LO}}$, $\mathbf{\Gamma}_{\mathrm{TO}}$, $\mathbf{K}$) were defined as the average  on the Fermi surface of the matrix element $\left|D_{\lambda \boldsymbol{k} \lambda{\prime}\boldsymbol{k}^{\prime}}^{\eta}\right|$ of the Kohn-Sham potential, differentiated with respect to the phonon displacement. The EPC for LO and TO modes at the $\Gamma$ point had $\left\langle D_{\bf{\Gamma}}^{2}\right\rangle_{\mathrm{F}}=45.6\,(\mathrm{eV\,\AA^{-1}})^2$, which was in good agreement with experimental results \cite{Basko2009a}. However, the EPC value at the $\mathbf{K}$ point has been debated \cite{Basko2009a, Ferrari2006b, Lazzeri2008, Berciaud2009b, Gruneis2009b} because it is renormalized by the e--e interaction and is affected by the presence of the substrate owing to the dielectric screening effect \cite{Basko2008e}.
	The amount calculated by DFT with generalized gradient approximation was $\left\langle D_{\textbf{K}}^{2}\right\rangle_{\mathrm{F}}=92.0\,(\mathrm{eV\AA^{-1}})^2$ \cite{Piscanec2004a}. However, a GW calculation, which considers the e--e interaction by approximating the self-energy using the product of the Green function and screened Coulomb potential, but neglects the vertex corrections, yielded $\left\langle D_{\textbf{K}}^{2}\right\rangle_{\mathrm{F}}=193\,(\mathrm{eV\,\AA^{-1}})^2$\cite{Onida2002, Lazzeri2008}.

	In this work, we investigate the hot carrier dynamics in photoexcited heavily doped graphene on a polyethylene terephthalate (PET) substrate using an OPTP and estimate the EPC strength at the $\mathbf{K}$ point via a numerical simulation based on the combination of BTE and comprehensive temperature model \cite{Yamashita2021}.
	Owing to the small change in the Drude weight of heavily doped graphene and negligible contribution of charged impurity and surface optical phonon (SOP) of PET substrate, the rise and relaxation dynamics of the SCOP are effectively captured by the transient THz reflectivity change measured by ultrabroadband THz probe. Using the renormalization group analysis, the obtained dimensionless EPC at $\mathbf{K}$ point is discussed and compared with those determined by other techniques.
		
	\section{Simulation method and results}
	In this section, we present a numerical simulation of the THz conductivity and the transient THz reflectivity measured by the OPTP experiment according to the following procedures. After photoexcitation, photoexcited carriers are quickly recombined and  their energy is redistributed within electron gas forming hot carrier state in quasi-equilibrium with single chemical potential.  A number of cooling pathways for hot carriers by inelastic scattering have been proposed such as SCOPs\cite{ Kampfrath2005,Mihnev2016a, Brida2013}, acoustic phonon\cite{Bistritzer2009, Song2012a, Graham2013, Betz2013}, SOP of substrate \cite{Low2012}. As we will explain later, the contribution of SOP and its coupled mode with plamons can be neglected by selecting the substrate with low polarizability and small phonon energy $\hbar \omega_{\mathrm{SO}}$ \cite{Bostwick2007b, Rana2011, Hamm2016a}.  Effect of acoustic phonon on hot carrier cooling is considered by the supercollision process and the acoustic phonon occupation is assumed to remain unchanged from the equilibrium state in the picosecond time scale after photoexcitation \cite{Johannsen2013b}. Therefore, we use comprehensive temperature model to calculate the temporal evolutions of the temperature for hot carriers in quasi-equilibrium and the occupations for three dominant SCOP modes. Thereafter, the iterative solution of BTE \cite{Yamashita2021} is used to calculate the intraband complex conductivity of the hot carriers in THz region. Because interband transition is forbidden at a THz probe energy of $\hbar \omega_{\mathrm{THz}} < 2 \varepsilon_{\mathrm{F}}$, the THz conductivity of doped graphene is dominated  by the intraband transition. This scheme enables us to reduce the computational cost substantially compared to the calculation of the full solutions of coupled graphene Bloch equation and BTEs for hot carriers and hot phonon modes in 2D momentum space. 
	
	\subsection{THz conductivity calculation}
	The iterative solution of the BTE for obtaining the steady-state and dynamical conductivity of semiconductors was introduced in Refs.\,\cite{Willardson1972a, Lundstrom2009a} and was subsequently modified for 2D MDF in graphene\cite{Yamashita2021}. 
	The temporal evolution of the carrier distribution is described by the BTE under a time-dependent electric field, which is expressed as 
	\begin{equation}
		\frac{\partial f_{\lambda}(\boldsymbol{k}, t)}{\partial t}=-\frac{(-e)}{\hbar} \boldsymbol{E}(t) \frac{\partial f_{\lambda}(\boldsymbol{k}, t)}{\partial \boldsymbol{k}}+
		\left.\frac{\partial f_{\lambda}(\boldsymbol{k}, t)}{\partial t}\right|_{\text {c }}.
	\end{equation}
	Here, $f_{\mathit{\lambda}}(\boldsymbol{k}, t)$ is the electron distribution function for the conduction band ($\lambda = 1$) and valence band ($\lambda =  -1$), $\boldsymbol{k}$ is the wave vector of the carriers, $e$ is the elementary charge, and $\boldsymbol{E}(t)$ is the electric field of the THz probe pulse. $\partial f_{\lambda}(\boldsymbol{k}, t) /\left.\partial t\right|_{\text {c }}$ is the collision term that describes the change in the distribution function via carrier scattering.

	We consider the intrinsic carrier scattering mechanism by the optical and acoustic phonon modes\cite{Tan2007b, Hwang2007k, Morozov2008b, Dean2010b, Perebeinos2010b, Tanabe2011a, Zou2010, Hwang2008g, Castro2010a, VanNguyen2020} and the extrinsic mechanism by the charged impurities \cite{Adam2007b, Tan2007b, Ando2006a, Chen2008h, Hwang2007k}, and weak scatterers \cite{Lin2014d, Stauber2007b, Adam2008b, Katsnelson2008a, Morozov2008b, Dean2010b, Pachoud2010b, Yan2011b}.
	For spherical bands under a low field $\boldsymbol{E}$, the general solution of Eq. (1) is approximately provided by the first two terms of the zone spherical expansion.
	\begin{equation}
		f_{\lambda}(\boldsymbol{k},t)=f_{0}\left(\varepsilon_{\lambda \boldsymbol{k}}\right)+g\left(\varepsilon_{\lambda \boldsymbol{k}},t\right) \cos \alpha_{\boldsymbol{k}},
	\end{equation}
	where $f_{0}\left(\varepsilon_{\lambda \boldsymbol{k}}\right)=1 /\left[\exp \left\{\left(\varepsilon_{\lambda \boldsymbol{k}}-\mu\left(T_{e}\right)\right) / k_{\mathrm{B}} T_{e}\right\}+1\right]$ is the Fermi-Dirac distribution for the corresponding equilibrium electron distribution at the electron temperature $T_{\mathrm{e}}$. $\varepsilon_{\lambda \boldsymbol{k}}=\pm \hbar v_{\mathrm{F}}|\boldsymbol{k}|$ ($\varepsilon_{1 \boldsymbol{k}} \geq 0$ and $\varepsilon_{-1 \boldsymbol{k}} \leq 0$ for the conduction and valence bands, respectively) is the electron energy within the Dirac approximation of the graphene energy-band structure \cite{Novoselov2009b}, and $v_{\mathrm{F}}$ is the Fermi velocity. In this expression, $\mu\left(T_{\mathrm{e}}\right)$ is the temperature-dependent chemical potential of the 2D MDF\cite{Ando2006a, Hwang2009e, Frenzel2014d} and is illustrated in Fig. 3(a). $g\left(\varepsilon_{\lambda \boldsymbol{k}}, t \right)$ is the perturbation part of the distribution, and $\alpha_{\boldsymbol{k}}$ is the angle between $\boldsymbol{E}$ and $\boldsymbol{k}$.
	
	In Eq. (1), the collision term is given by
	\begin{equation}
		\left.\frac{\partial f_{\lambda}(\boldsymbol{k}, t)}{\partial t}\right|_{\text {c }}=\sum_{\eta, \lambda^{\prime}} C_{\lambda \lambda^{\prime}}^{\eta}(\boldsymbol{k})+C_{\lambda}^{\mathrm{el}}(\boldsymbol{k}),\end{equation}
	while accounting for the scattering of the electrons with dominant optical phonon modes $\eta$, in $C_{\lambda \lambda}^{\eta}$, including both the intraband $\left(\lambda=\lambda^{\prime}\right)$ and interband $\left(\lambda \neq \lambda^{\prime}\right)$ processes with elastic scattering processes in $C_{\lambda}^{\mathrm{el}}(\boldsymbol{k}).$ The carrier collision term $C_{\lambda \lambda^{\prime}}^{\eta}(\boldsymbol{k})$ for the interaction of the electron and optical phonons is expressed as:
	\begin{equation}
		\begin{aligned}
			C_{\lambda \lambda^{\prime}}^{\eta}(\boldsymbol{k})=\sum_{\mathbf{k^{\prime}}}&\left\lbrace P^{\eta}_{\lambda^{\prime}\mathbf{k}^{\prime}  \lambda\mathbf{k} } f_{\lambda^{\prime}} (\mathbf{k^{\prime}})(1-f_{\lambda} (\mathbf{k}))\right.\\
			&\left.-P^{\eta}_{\lambda\mathbf{k}  \lambda^{\prime} \mathbf{k^{\prime}} } f_{\lambda} (\mathbf{k})(1-f_{\lambda^{\prime}} (\mathbf{k}^{\prime}))	\right\rbrace
		\end{aligned}
	\end{equation}
	where $P^{\eta}_{\lambda^{\prime} \mathbf{k}^{\prime}  \lambda\mathbf{k} }$ and $P^{\eta}_{\lambda \mathbf{k}  \lambda^{\prime}\mathbf{k}^{\prime} }$ are the transition rate by the optical phonon modes, $\eta$, between states $(\mathbf{k}^{\prime}, \lambda^{\prime}) \to (\mathbf{k}, \lambda)$ and $(\mathbf{k}, \lambda) \to (\mathbf{k}^{\prime}, \lambda^{\prime})$, respectively. $P^{\eta}_{\mathbf{k} \lambda \mathbf{k}^{\prime} \lambda^{\prime}}$ is expressed 
	by
	\begin{equation}
		P_{ \lambda \boldsymbol{k}  \lambda^{\prime}\boldsymbol{k}^{\prime}}^{\eta}=P_{ \lambda \boldsymbol{k}  \lambda^{\prime}\boldsymbol{k}^{\prime}}^{\mathrm{EM}, \eta}+P_{ \lambda \boldsymbol{k}  \lambda^{\prime}\boldsymbol{k}^{\prime}}^{\mathrm{AB}, \eta},
	\end{equation}
	which accounts for the phonon emission and absorption, given by
	\begin{equation}\begin{aligned}
			P_{ \lambda \boldsymbol{k}  \lambda^{\prime}\boldsymbol{k}^{\prime}}^{\mathrm{EM/AB}, \eta} =&\frac{\pi\left|D_{\lambda\boldsymbol{k} \lambda^{\prime} \boldsymbol{k}^{\prime}}^{\eta}\right|^{2}}{\rho \omega_{\eta} A}\left(n_{\eta}+\frac{1}{2} \pm \frac{1}{2}\right) \\
			& \times \delta\left(\varepsilon_{\lambda \boldsymbol{k}}-\varepsilon_{\lambda^{\prime} \boldsymbol{k}^{\prime}} \mp \hbar \omega_{\eta}\right) \delta\left(\boldsymbol{k}-\boldsymbol{k}^{\prime} \mp \boldsymbol{q}\right).
	\end{aligned}\end{equation}
	Here, $\left|D_{\lambda \boldsymbol{k} \lambda{\prime}\boldsymbol{k}^{\prime}}^{\eta}\right|$ is the EPC matrix element defined in Ref.\,\cite{Piscanec2004a}, $\boldsymbol{k}^{\prime}=\boldsymbol{k} \pm \boldsymbol{q}$, and $\boldsymbol{q}$ is the wave vector of the optical phonons. $\rho = 7.6 \times 10^{-7}\,\mathrm{kgm}^{-2}$ is the area density of graphene and $A$ is the area of the graphene sample, whereas $\omega_{\eta}$ and $n_{\eta}$ are the angular frequency and occupation of the optical phonons, respectively. 
	
	The carrier-scattering rates that are obtained by the optical phonons in Eq.\;(6) account for the phonon emission and absorption. For small $\boldsymbol{q}$ and $\boldsymbol{k}^{\prime}$, the EPC matrix elements $|D_{\lambda \bm{k} \lambda^{\prime} \bm{k}^{\prime}}^{\eta}|^{2}$ for the $\mathbf{\Gamma}_{\mathrm{LO}}$, $\mathbf{\Gamma}_{\mathrm{TO}}$, and $\mathbf{K}$ phonons are expressed by \cite{Piscanec2004a, Piscanec2007c}
	\begin{equation}
		\begin{aligned}
			\left|D_{\lambda \bm{k} \lambda^{\prime}\bm{k}^{\prime}}^{\mathbf{\Gamma}_{\mathrm{LO/TO}}}\right|^{2}&=\left\langle D_{\mathbf{\Gamma}}^{2}\right\rangle_{\mathrm{F}}\left\{1 \pm \cos \left(\theta_{\bm{k},\bm{q}}+\theta_{\bm{k}^{\prime}, \bm{q}}\right)\right\}, \\
			\left|D_{\lambda \bm{k} \lambda^{\prime}\bm{k}^{\prime}}^{\mathbf{K}}\right|^{2}&=\left\langle D_{\mathrm{\textbf{K}}}^{2}\right\rangle_{\mathrm{F}}\left\{1 \pm \cos \theta_{\bm{k},\bm{k}^{\prime}} \right\}.
		\end{aligned}
	\end{equation}
	Here, $\theta_{\bm{k},\bm{q}}$ denotes the angle between $\boldsymbol{k}$ and $\boldsymbol{q}$, $\theta_{\bm{k}^{\prime},\bm{q}}$ denotes the angle between $\boldsymbol{k}^{\prime}$ and $\boldsymbol{q}$, and $\theta_{\bm{k},\bm{k}^{\prime}}$ denotes the angle between $\boldsymbol{k}$ and $\boldsymbol{k^{\prime}}$. In the case of $\mathbf{\Gamma}_{\mathrm{LO}}$ and $\mathrm{\bf{K}}$ phonons, the plus sign refers to the interband processes, and for $\mathbf{\Gamma}_{\mathrm{TO}}$ phonons, it refers to the intraband processes. 
	
	In Eq. (3), the elastic term $C_{\lambda}^{\mathrm{el}}(\boldsymbol{k})$ is calculated using the elastic scattering rate $P_{\lambda \bm{k} \lambda \bm{k}^{\prime}}^{s}$\cite{Yamashita2021}. The index, $s$, refers to the different elastic scattering modes by weak scatterers, and charged impurities, which are characterized by resistivity of the weak scatterers $\rho_{\mathrm{s}}$, and charged impurity concentration $n_{\mathrm{i}}$, respectively. The reported  $\rho_{\mathrm{s}}$ ranges from 40--100 $\Omega$\cite{Morozov2008b, Dean2010b, Pachoud2010b, Yan2011b}. Interactions with acoustic phonons are treated in a quasi-elastic and included in $C_{\lambda}^{\mathrm{el}}(\boldsymbol{k})$. Different electron-acoustic phonon coupling models have been proposed to extract the effective coupling constant $J_a$ from experimental data for graphene which ranges 10-30 eV\cite{ Stauber2007b, Bolotin2008b, Chen2008b, Hwang2008g, Hong2009a, Zou2010, Dean2010b,  Efetov2010a, Castro2010a, Perebeinos2010, Kozikov2010, Mariani2010a, Min2011, Kaasbjerg2012a, Ochoa2012, Sohier2014}. A first-principle study suggests that the gauge-field contribution is more important than the screened deformation potential\cite{Borysenko2010b, Park2014}.
	
	The iterative solution of $g^{j}\left(\varepsilon_{\lambda \textbf{\textit{k}}}\right)=g\left(\varepsilon_{\lambda \textbf{\textit{k}}}, t_j\right)$  is provided by
	\begin{equation}\begin{aligned}
			g^{j+1}&\left(\varepsilon_{\lambda \textbf{\textit{k}}}\right)=\cfrac{S_{\lambda}^{\mathrm{in}}-\cfrac{(-e)E^j}{\hbar} \cfrac{\partial f_{0}}{\partial k}+\Omega_{\mathrm{s}} g^{j}}{S_{\lambda}^{\mathrm{out}}+\nu^{\mathrm{el}}+\Omega_{\mathrm{s}}}.
	\end{aligned}\end{equation}
	Here, $E^j=|\textbf{\textit{E}}(t_{j})|$ and $\textit{k}=|\textbf{\textit{k}}|$ are the magnitudes of the electric field and wavevector, respectively. $\Omega_{\mathrm{s}}$ is known as the self-scattering rate, and $1/\Omega_{\mathrm{s}}$ is the time increment between successive iterations, and $S_{\lambda}^{\mathrm{in}}$ and $S_{\lambda}^{\mathrm{out}}$ are the net in- and out-scattering rates for inelastic scattering, respectively. Furthermore, $\nu^{\mathrm{el}}$ is the total relaxation rate by the elastic scattering mechanisms. The sequence $\{g^{j}\left(\varepsilon_{\lambda \textbf{\textit{k}}}\right)\}$ yields $f_{\lambda}(\textbf{\textit{k}}, t_j)$ versus time when $\Omega_{\mathrm{s}}$ is sufficiently large compared to $S_{\lambda}^{\text {out }}+\nu^{\text {e }}$. 
	
	\subsection{Temperature model of hot carriers}
	The hot carrier intraband optical conductivity $\sigma (\omega, \tau_1)$ in the cooling process can be calculated from $f_{\lambda}\left(\textbf{\textit{k}}, t_j\right)$, which is obtained by substituting the hot carrier and three dominant optical phonon temperatures ($T_e (t_j)$, $T_\eta (t_j)$) into Eq. (8) in the iteration process. Here, $\tau_1$ is the pump probe delay. We employ the coupled rate equations for a comprehensive temperature model that  describe the temperature evolutions of the electron temperature  $T_{e}$  and optical phonon occupations $n_{\eta}$ by photoexcitation: 
	\begin{subequations}
		\begin{equation}
			\frac{d T_{e}}{d t}=\frac{\mathcal{I}_{\mathrm{ab}}-\sum_{\eta} R_{\eta}^{\mathrm{Net}} \hbar \omega_{\eta}-J_{\mathrm{sc}}}{C}\\,
		\end{equation}
		\begin{equation}
			\frac{d n_{\eta}}{d t}=R_{\mathrm{M}, \eta}^{\mathrm{Net}}-\frac{n_{\eta}-n_{\eta 0}}{\tau_{\mathrm{ph}}}.
		\end{equation}
	\end{subequations}
	In this case, $\mathcal{I}_{\mathrm{ab}}$ represents the pump intensity absorbed in graphene sample during laser irradiation, considering the multiple reflections inside the substrate with dielectric constant $\epsilon (\omega_{\mathrm{pump}})$=2.4 for the pump wavelength and saturable absorption (SA) effect in graphene. $C$ is the sum of the specific heat of the electrons in the conduction and valence bands, $R_{\eta}^{\mathrm{Net}}=R_{\eta}-G_{\eta}$ denotes the total balance between the optical phonon emission and absorption rate, and $J_{\mathrm{sc}}$ indicates the energy loss rate for the supercollision carrier-cooling process \cite{Song2012a, Someya2017b}. $R_{M, \eta}^{\text {Net }}=R_{\mathrm{M}, \eta}-G_{\mathrm{M}, \eta}$ denotes the total balance between the optical phonon emission and absorption rate per number of phonon modes that participate the carrier scattering. In calculations of $R_{\eta}^{\mathrm{Net}}$ and $R_{M, \eta}^{\text {Net }}$, we include the scattering angle dependence of the $|D_{\lambda \boldsymbol{k}, \lambda^{\prime} \bm{k}^{\prime}}^{\eta}|^{2}$ in Eq.(7) which have not been considered in the temperature model used in the previous study \cite{Rana2009c, Wang2010a, Someya2017b, Yamashita2021}. Moreover, $n_{\eta0}$ represents the phonon occupation near the $\mathbf{\Gamma}$ and $\mathbf{K}$ points, respectively, in equilibrium at room temperature, whereas $\tau_{\mathrm{ph}}$ is the phenomenological optical phonon decay time to other phonon modes 
	via the phonon--phonon interaction caused by lattice anharmonicity \cite{Bonini2007}. The effective optical phonon temperatures are calculated by inverting the Bose--Einstein distribution function, $n_{\eta}=1 /(e^{\hbar \omega_{\eta} / k_{\mathrm{B}} T_{\eta}}-1)$. The formula and temperature dependence of  $C$, $R_{\eta}^{\mathrm{Net}}$, $J_{\mathrm{sc}}$ and $R_{M, \eta}^{\text {Net }}$ can be found in Ref. \cite{Yamashita2021} and Section SIII in Supplemental Material (SM).
	
	\begin{table}[b]
		\caption{\label{tab:fonts} Parameters of graphene on PET substrate and experimental setups used in simulation. The dielectric properties of the PET substrate were obtained from Refs \cite{Fedulova2012, Zhang2020}.}
		\label{table:data_type}
		\centering
		\begin{tabular}{ccc}
			\hline
			Quantity& Lightly doped & Heavily doped \\
			\hline\hline $|\varepsilon_{\mathrm{F}}|\,(\mathrm{eV})$ & 0.15 & 0.43 \\
			$v_{\mathrm{F}}\,(\mathrm{m\,s}^{-1})$ & \multicolumn{2}{c}{ $1.1 \times 10^{6}$} \\
			$\epsilon_{\mathrm{s}}$ & \multicolumn{2}{c}{3.0}\\			
			$\epsilon (\mathrm{\omega_{THz}})$ & \multicolumn{2}{c}{2.5} \\
			$\epsilon (\mathrm{\omega_{pump}})$ & \multicolumn{2}{c}{2.4} \\
			$J_{\mathrm{a}}\,(\mathrm{eV})\textsuperscript{\emph{a}}$ & 30, 20 & 30, 20\\
			$\left\langle D_{\mathbf{\Gamma}}^{2}\right\rangle_{\mathrm{F}}\,(\mathrm{eV}\,\mathrm{\AA}^{-1})^{2}$& \multicolumn{2}{c}{45.6}\\
			$\left\langle D_{\mathbf{K}}^{2}\right\rangle_{\mathrm{F}}\,(\mathrm{eV}\,\mathrm{\AA}^{-1})^{2}$& \multicolumn{2}{c}{92.0--703}\\
			$\rho_{\mathrm{s}}\,(\Omega)$& \multicolumn{2}{c}{100}\\
			$n_{\mathrm{i}} (10^{12}\,\mathrm{cm^{-2}})$\textsuperscript{\emph{a}}&0, 0.17 & 0, 1.7\\
			$\sigma_{\mathrm{DC}} (G_0)$& 25.5, 25.5 & 25.7, 25.9\\
			$\tau_{\mathrm{ph}}\,(\mathrm{ps})$& \multicolumn{2}{c}{1.0}\\
			$\tau_{\mathrm{ie}}\,(\mathrm{fs})$& \multicolumn{2}{c}{100}\\
			$F_0 (\mathrm{\mu J\,cm^{-2}})$ & \multicolumn{2}{c}{100}\\
			$2\tau_{\mathrm{pump}}\,(\mathrm{fs})$& \multicolumn{2}{c}{220}\\
			$2\tau_{\mathrm{prob}}\,(\mathrm{fs})$& \multicolumn{2}{c}{300}\\
			\hline
		\end{tabular}\\
		\textsuperscript{\emph{a}} The $D_{\mathrm{ac}}$ and the $n_{\mathrm{i}}$ values were chosen to give the nearly equal DC conductivity $\sigma_{\mathrm{DC}}$.
	\end{table}

	The  optical pump pulse is absorbed by interband transition and the absorption coefficient for free-standing graphene at the normal incidence is $\alpha_{\mathrm{inter}}=\pi \alpha=0.23\,\%$ under sufficiently weak pump condition, where $\alpha$ is the fine structure constant. However, the SA effect in graphene under the intense pump fluence\cite{Hasan2009, Bao2009, Xing2010, Marini2017a} should be considered. The SA is a nonperturbative, nonlinear optical phenomenon that depends on the pump power as well as the temperature and Fermi energy. Based on the theory by Marini et al.\cite{Marini2017a}, we derived the formula of $\mathcal{I}_{\mathrm{ab}}$ considering the SA and multiple reflections inside the substrate at the oblique angle of incidence for the temperature calculation in the experimental condition (see Section IV in the {SM):
	\begin{equation}
		\begin{aligned}
			\mathcal{I}_{ab}(t)=\mathcal{I}_0(t) A^{s*}_{\mathrm{12}}+\sum \mathcal{I}_n(t+ n \Delta T) A^{s*}_{\mathrm{21}},
		\end{aligned}
	\end{equation}
	where $\mathcal{I}_0(t)$ is the envelope function of the incident pump pulse, which is assumed to have hyperbolic secant form, $\mathcal{I}_0(t)=\left(F_{\mathrm{0}} / 2 \tau_{\mathrm{pump}}\right) \operatorname{sech}^{2}\left(t / \tau_{\mathrm{pump}}\right)$. In this case, $F_{\mathrm{0}}$ is the incident fluence and $ 2 \tau_{\mathrm{pump}}$ is the pump pulse duration. $\mathcal{I}_n(t+n\Delta T)=\left(F_{n} / 2 \tau_{\mathrm{pump}}\right) \operatorname{sech}^{2}\left((t+n \Delta T) / \tau_{\mathrm{pump}}\right)$ represents the pump pulse by the n-th multiple reflection of the incident pump pulse inside the substrate, where $F_{\mathrm{n}}$ is the fluence and $n \Delta T$ is the round-trip time for the n-th reflection pump pulse in the substrate. $A^{s*}_{\mathrm{ij}}(F_{\mathrm{0}} / 2 \tau_{\mathrm{pump}})$ is the absorption coefficient including the carrier temperature dependence of the SA effect at the interface of layer i/graphene/layer j when the pump pulse excites the graphene from layer i (see Fig. S1 of Section I in the SM). In this model, the SA is characterized by the inelastic carrier relaxation time $\tau_{\mathrm{ie}}$.
	The pump intensity dependence of the interband absorption coefficient $\alpha_{\mathrm{inter}}$ for the free standing graphene and $A^{s*}_{\mathrm{ij}}(F_{\mathrm{0}} / 2 \tau_{\mathrm{pump}})$ for the graphene on substrate can be seen in Figs.S2 and S3 of Section IV in the SM.
	
	\subsection{Simualtion for graphene on PET substrate }
	In the simulation, the carrier scattering by SOPs of substrate are not included, while the SOPs play crucial roles for the carrier dynamics in graphene on polar substrate \cite{Chen2008b, Fratini2008a, Li2010b, Konar2010b, Hwang2013e, Tielrooij2018}. The square of EPC matrix element between SOP and carries is proportional to 
	\begin{equation}
			g_{\mathrm{SO}} \frac{e^{-qd}}{q+q_s}
	\end{equation}
	Here, $g_{\mathrm{SO}} =\beta e^2 \hbar \omega_{\mathrm{SO}}/2 \epsilon_0$, $\omega_{\mathrm{SO}}$ is the angular frequency of the SOP, $\varepsilon_0$ is permittivity of vacuum and $d$ is the equilibrium distance of the graphene sheet from the substrate surface. $q$ is the angular wavenumber of the surface phonon, $q_s$ is the Thomas-Fermi screening constant of the 2D carriers and 
	\begin{equation}
		\beta=\frac{\epsilon_{\mathrm{s}}-\epsilon_{\infty}}{(\epsilon_{\mathrm{s}}+1)(\epsilon_{\mathrm{\infty}}+1)},
	\end{equation}
	where $\epsilon_{\mathrm{s}}$ and  $\epsilon_{\mathrm{\infty}}$ are the low and high frequency dielectric constant, respectively.  $\beta$ is a measure of the polarizability of the dielectric interface. 
	
	For example, in crystalline $\mathrm{SiO_2}$ ($\epsilon_{\mathrm{s}}=3.9$, $\epsilon_{\mathrm{\infty}}=2.5$) has two SOP modes  at $\hbar \omega_{\mathrm{s1}}=60.0\,\mathrm{meV}$, $\hbar \omega_{\mathrm{s2}}=146.5\,\mathrm{meV}$, with $\beta_1=0.025$ $\beta_2=0.062$, respectively. These values correspond to $g_{\mathrm{SO}1}=0.14\,(\mathrm{eV^2\,\AA^{-1}})$ and $g_{\mathrm{SO}2}=0.82\,(\mathrm{eV^2\,\AA^{-1}})$ and are enhanced by roughly 50 \% in conventional $\mathrm{SiO_2}$ glass with $\epsilon_{\mathrm{\infty}}=2.1$. As a result, the temperature dependence of carrier transport is dominated by SOP scattering in graphene on polar substrate such as $\mathrm{SiO_2}$ and  $\mathrm{HfO_2}$ \cite{Chen2008b, Konar2010b}. The energy loss rate of hot carrier by SOP modes is given as  $R_{\mathrm{SO}}^{\mathrm{NET}} \propto g_{\mathrm{SO}} \hbar \omega_{\mathrm{SO}}$ so that the large $\hbar \omega_{\mathrm{SO}}$ also affect the hot carrier dynamics significantly. The dispersion relation of SOP modes can be altered by the coupling of plasmon and SOP in doped graphene. These effects change significantly the hot carrier dynamics and makes the simulation more complex leading to hindering the estimation of EPC at K point. 
	
	Therefore, in this study, we select graphene sample on a PET substrate which has the low polarizability ($\epsilon_{\mathrm{s}}=3.0$, $\epsilon_{\mathrm{\infty}}=2.54$) owing to the polar low frequency vibrational modes around $10\,\mathrm{meV}$ \cite{Fedulova2012}. The $g_{\mathrm{SO}}=0.029 \,(\mathrm{eV^2 \AA^{-1}})$ of PET is small and decreases significantly in doped graphene by carrier screening effect. The $R_{\mathrm{SO}}^{\mathrm{NET}}$ between carriers and SOP of PET is expected to be smaller by 3 orders of magnitude than $\mathrm{SiO_2}$ and makes the negligible contribution on hot carrier cooling and THz conductivity. Furthermore, the small static dielectric constant $\epsilon_{\mathrm{s}}=3.0$ of a PET substrate provides weak dielectric screening with an expected larger renormalization effect on the EPC by e--e interaction\cite{Basko2008e}.

	The transient reflection change $\Delta E_{\mathrm{r}}(\tau_1)/E_0$ of graphene on PET substrate with  the dielectric constant $\epsilon (\omega_{\mathrm{THz}})=2.5$ in THz region can be calculated from the $\sigma (\omega, \tau_1)$. (For details, see Section V in the SM). In this case, $\Delta E_{\mathrm{r}}(\tau_1)/E_0$ is defined as $\Delta E_{\mathrm{r}}(\tau_1)/E_0 \equiv \left(E_{\mathrm{r}}\left(\tau_{2}, \tau_{1}\right)-E_{\mathrm{r}}\left(\tau_{2}\right)\right) / E_{\mathrm{r}}\left(\tau_{2}\right)$ at the probe trigger delay $\tau_{2}=0\,\mathrm{ps}$ when the electric field of the THz probe pulse exhibits the maximum amplitude as seen in Fig. 5. $E_{\mathrm{r}}\left(\tau_{2}, \tau_{1}\right)$ and $E_{\mathrm{r}}\left(\tau_{2}\right)$ are the THz electric fields that are reflected  from the graphene with and without photoexcitation, respectively. $\Delta E_{\mathrm{r}}\left(\tau_{1}\right) / E_{0}$ is useful for discussing the hot carrier relaxation and photoconductivity, $\Delta \sigma\left(\omega, \tau_{1}\right)=\sigma\left(\omega, \tau_{1}\right)-\sigma_{0}(\omega)$, around the center frequency of the THz probe pulse, where $\sigma_{0}(\omega)$ is the intraband optical conductivity of graphene without pump fluence. 
	$\Delta E_{\mathrm{r}}\left(\tau_{1}\right) / E_{0}>0$ and $\Delta E_{\mathrm{r}}\left(\tau_{1}\right) / E_{0}<0$ indicate the positive and negative photoconductivities, $\Delta \sigma_{1}\left(\omega, \tau_{1}\right)$, respectively.
				
	We investigated the effect of the EPC on the hot carrier dynamics of photoexcited graphene on the PET substrate for different Fermi energies $\varepsilon_{\mathrm{F}}$. The parameters used in the simulation are summarized in Table I.
	Figures 1(a) and (b) depict the temporal evolutions of $T_e$ and $T_{\eta}$ in the heavily doped graphene with $|\varepsilon_{\mathrm{F}}|=0.43\,\mathrm{eV}$ for $\left\langle D_{\textbf{K}}^{2}\right\rangle_{\mathrm{F}}=193$ and $703 \,(\mathrm{eV\,\AA^{-1}})^2$ under the pump fluence $F_0=100 \,\mathrm{\mu J\,cm^{-2}}$ calculated using the temperature model. In this case, $\left\langle D_{\mathbf{\Gamma}}^{2}\right\rangle_{\mathrm{F}}$ is fixed at the DFT value because the EPC of the $\mathbf{\Gamma}_{\mathrm{LO/TO}}$ phonon is not affected by the e-e interaction and well agree with the experiment \cite{Basko2009a}. The difference of $T_{\Gamma_{\mathrm{LO}}}$ and $T_{\Gamma_{\mathrm{TO}}}$ stems from the scattering angle dependence of $\left|D_{\lambda \bm{k} \lambda^{\prime}\bm{k}^{\prime}}^{\mathbf{\Gamma}_{\mathrm{LO/TO}}}\right|^{2}$ in Eq. (7). 
	A comparison between Figs.1 (a) and (b) reveals that the rise and relaxation dynamics of the hot carrier and optical phonon temperatures depend significantly on $\left\langle D_{\textbf{K}}^{2}\right\rangle_{\mathrm{F}}$. At $\left\langle D_{\textbf{K}}^{2}\right\rangle_{\mathrm{F}}=703 \,(\mathrm{eV\,\AA^{-1}})^2$, $T_{\mathbf{K}}$ followed $T_e$ more rapidly and increases up to 1800 K much higher than $T_{\mathbf{\Gamma}_{\mathrm{LO/TO}}}$, indicating that substantially more hot carrier energy is mainly transferred  into the $\mathbf{K}$  phonon owing to the stronger EPC. As a result, the maximum $T_e$ for $\left\langle D_{\textbf{K}}^{2}\right\rangle_{\mathrm{F}}=703 \,(\mathrm{eV\,\AA^{-1}})^2$ becomes lower than that for $\left\langle D_{\textbf{K}}^{2}\right\rangle_{\mathrm{F}}=193 \,(\mathrm{eV\,\AA^{-1}})^2$.
	\begin{figure} [t]
		\centering
		\includegraphics[width=8.5cm, bb=0 0 340 227] {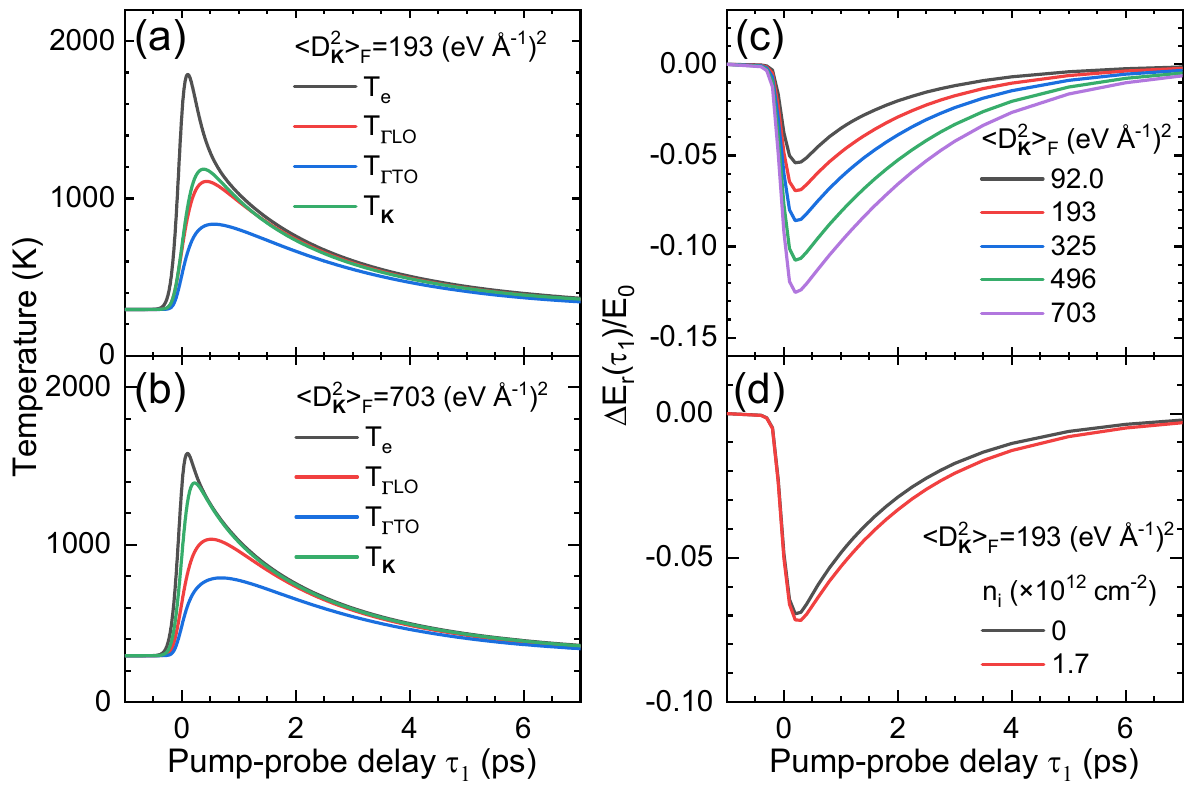}
		\caption{Simulation results of heavily doped graphene with $|\varepsilon_F|=0.43\,{\mathrm{eV}}$ for $F_0=100\,\mathrm{\mu J\,cm^{-2}}$.  Temporal evolutions of $T_{\mathrm{e}}$ and $T_{\eta}$  for $\left\langle D_{\textbf{K}}^{2}\right\rangle_{\mathrm{F}}$=(a) 193 and (b) 703 $(\mathrm{eV\,\AA^{-1})^2}$. (c) $\left\langle D_{\textbf{K}}^{2}\right\rangle_{\mathrm{F}}$ dependence of  $\Delta E_{\mathrm{r}}\left(\tau_{1}\right) / E_{0}$ of graphene calculated using temporal waveforms of THz probe pulse expressed by second derivative of Gaussian function, $\exp (-t^{2} / \tau_{\mathrm{pump}}^{2})$, with pulse durations of $2\tau_{\mathrm{pump}}=300\,\mathrm{fs}$.
		(d) $\Delta E_{\mathrm{r}}\left(\tau_{1}\right) / E_{0}$ at $\left\langle D_{\textbf{K}}^{2}\right\rangle_{\mathrm{F}}=193\,(\mathrm{eV\,\AA^{-1})^2}$ for $n_{\mathrm{i}}=0$ and $1.7 \times10^{12}\,\mathrm{cm}^{-2}$.}
		\label{fgr:example}
	\end{figure}	
	\begin{figure} [t]
		\centering
		\includegraphics[width=8.5cm, bb=0	0 340 226] {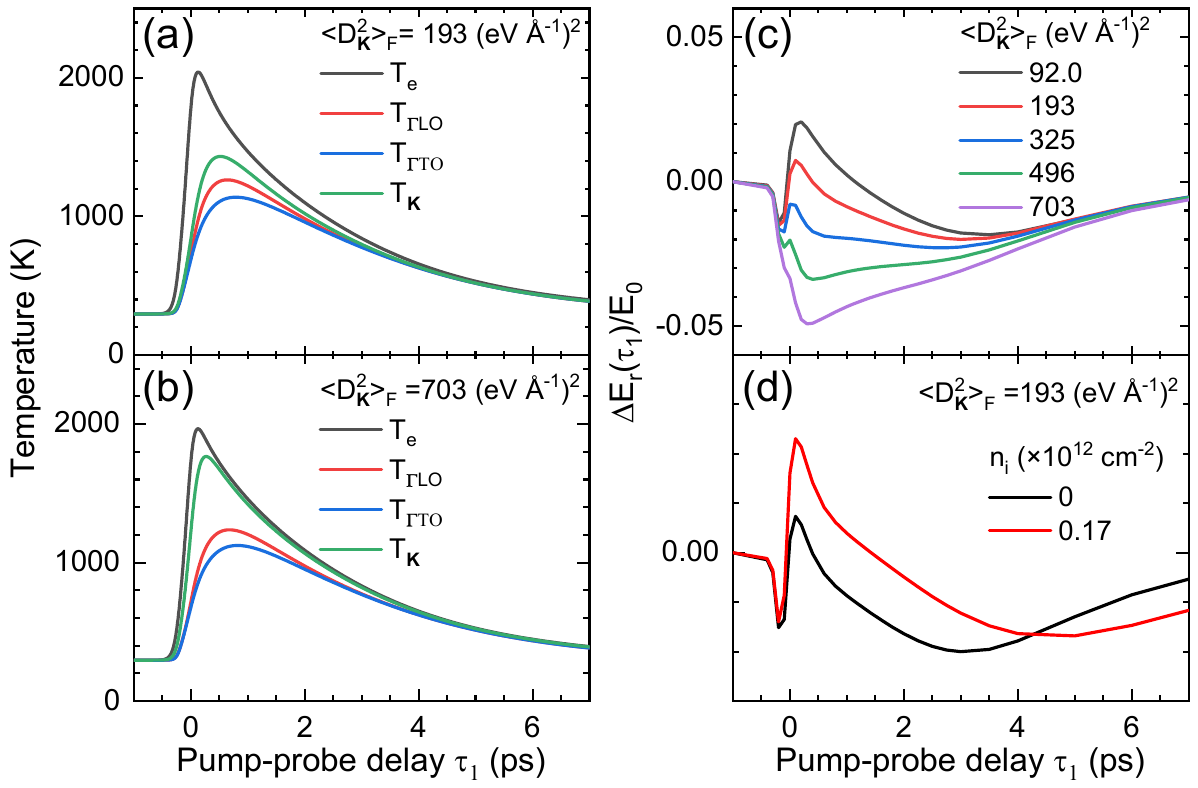}
		\caption{Simulation results of lightly doped graphene with $|\varepsilon_F|=0.15\,{\mathrm{eV}}$. Temporal evolutions of  $T_{\mathrm{e}}$ and $T_{\eta}$ for $\left\langle D_{\textbf{K}}^{2}\right\rangle_{\mathrm{F}}$=(a) 92.0 and (b) 703 $(\mathrm{eV\,\AA^{-1})^2}$ for $F_0=100\,\mathrm{\mu J\,cm^{-2}}$. (c) $\left\langle D_{\textbf{K}}^{2}\right\rangle_{\mathrm{F}}$ dependence of  $\Delta E_{\mathrm{r}}\left(\tau_{1}\right) / E_{0}$ of graphene. (d) $\Delta E_{\mathrm{r}}\left(\tau_{1}\right) / E_{0}$ at $\left\langle D_{\textbf{K}}^{2}\right\rangle_{\mathrm{F}}=193\,(\mathrm{eV\,\AA^{-1})^2}$ for $n_{\mathrm{i}}=0$ and $0.17\times10^{12}\,\mathrm{cm}^{-2}$.}
		\label{fgr:example}
	\end{figure}	
	Figure 1(c) presents the $\left\langle D_{\textbf{K}}^{2}\right\rangle_{\mathrm{F}}$ dependence of the transient reflection change $\Delta E_{\mathrm{r}}(\tau_1)/E_0$ calculated from the $\sigma (\omega, \tau_1)$ using the THz probe pulse with $2\tau_{\mathrm{p}}=300\,\mathrm{fs}$.
	The sign of $\Delta E_{\mathrm{r}}(\tau_1)/E_0$ remains negative indicating the negative photoconductivity as varying the $\left\langle D_{\textbf{K}}^{2}\right\rangle_{\mathrm{F}}$. The peak value of $\left| \Delta E_{\mathrm{r}}(\tau_1)/E_0 \right| $ increases monotonically as $\left\langle D_{\textbf{K}}^{2}\right\rangle_{\mathrm{F}}$ increases and effectively reflects the enhancement of $T_{\mathbf{K}}$.
	
	 Figure 2 depicts the simulation results on the lightly doped graphene with $|\varepsilon_{\mathrm{F}}|=0.15\,\mathrm{eV}$. Although the same phonon decay time $\tau_{\mathrm{ph}}=1\,\mathrm{ps}$ is used, the relaxation time of $T_e$ of the lightly doped graphene is longer than that of the heavily doped graphene owing to the weaker $R^{\mathrm{Net}}_{\eta}$ originated from the small density of state at the Fermi energy $\varepsilon_{\mathrm{F}}$. The sign of $\Delta E_{\mathrm{r}}(\tau_1)/E_0$ indicated in Fig. 2(c) changes depending on $\left\langle D_{\textbf{K}}^{2}\right\rangle_{\mathrm{F}}$ in contrast with the heavily doped graphene. For a small $\left\langle D_{\textbf{K}}^{2}\right\rangle_{\mathrm{F}}=92.0\,(\mathrm{eV\AA^{-1})^2}$,  $\Delta E_{\mathrm{r}}(\tau_1)/E_0$ 
	 exhibits positive photoconductivity, which is transformed into negative photoconductivity as $\left\langle D_{\textbf{K}}^{2}\right\rangle_{\mathrm{F}}$ increases. 
	
	The different behaviors in $\Delta E_{\mathrm{r}}(\tau_1)/E_0$ between the heavily and lightly doped graphene can be understood by considering the temperature dependence of the Drude weight $D\left(T_{\mathrm{e}}\right)$ of the graphene 2D MDF, which is the oscillator strength of free carrier absorption and plays a crucial role in carrier screening.  As can be observed in Fig.\,3(a), the chemical potential $\mu (T_e)$ of graphene 2D MDF decreases with $T_e$, leading to the unique temperature dependence of $D\left(T_{\mathrm{e}}\right)$ according to $\varepsilon_{\mathrm{F}}$ \cite{Muller2009, Gusynin2009, Wagner2014, Frenzel2014d, Yamashita2021}. In the case of a constant carrier relaxation rate, $D\left(T_{\mathrm{e}}\right)$ is expressed as
	\begin{equation}
	D\left(T_e\right)=\frac{2 e^{2}}{\hbar^{2}} k_{\mathrm{B}} T_{e} \ln \left[2 \cosh \left(\frac{\mu\left(T_{e}\right)}{2 k_{\mathrm{B}} T_{e}}\right)\right].
	\end{equation}
 	The $D(T_e)$ of the undoped graphene with $|\varepsilon_{\mathrm{F}}|=0.01\,\mathrm{eV}$ in Fig.\,3(b) increases linearly with $T_e$, yielding positive photoconductivity. However, $D(T_e)$ of the heavily doped graphene with $|\varepsilon_{\mathrm{F}}|=0.43\,\mathrm{eV}$ decreases slightly as $T_e$ increases and exhibits the minimum at around $T_e=2000 \,\mathrm{K}$, contributing to the negative photoconductivity below $T_e=3000\,\mathrm{K}$. At temperatures below 3000\,K, the maximum change in $D(T_e)$ is only 13$\%$ and the temperature dependence of THz conductivity change is mainly dominated by of the carrier scattering with the SCOPs. In the lightly doped graphene, $D(T_e)$ increases significantly above $T_e=1000\;\mathrm{K}$ and the contributions of $D(T_e)$ and the carrier scattering with SCOPs to the photoconductivity compete with one another resulting in the positive and negative photoconductivity depending on $T_e$ and $\left\langle D_{\textbf{K}}^{2}\right\rangle_{\mathrm{F}}$. 
 	\begin{figure}[t]
 		\centering
 		\includegraphics[width=8.5cm, bb=0 0 327 127] {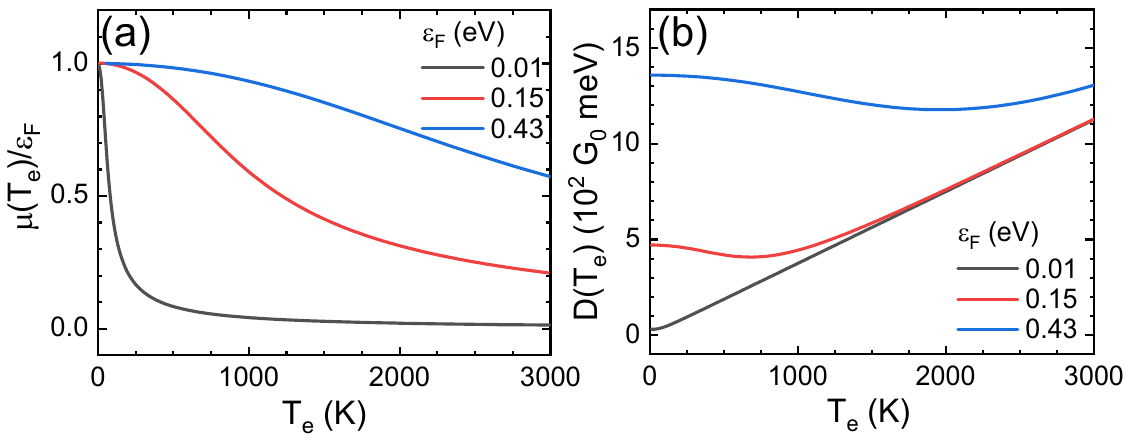}
 		\caption{ (a) $T_e$ dependence of chemical potential $\mu (T_e)$ and Drude weight $D(T_e)$ of graphene with $\varepsilon_{\mathrm{F}}=0.01, 0.15,$ and $0.43\,\mathrm{eV}$.}
 		\label{fgr:example}
 	\end{figure}
 	
 	We also investigated the effect of the charged impurity on the hot carrier dynamics in the heavily and lightly doped graphene because the charged impurity is one of the dominant scattering mechanism in graphene on substrate \cite{Tan2007b, Chen2008b,Adam2007b, Hwang2007k}.
 	Figure 1(d) shows the $\Delta E_{\mathrm{r}}(\tau_1)/E_0$ of the heavily doped graphene is almost unaffected by charged impurity scattering owing to the strong carrier screening effect. Here, the effective coupling constant  $J_a$ of acoustic phonon is selected so that the DC conductivity is almost equal as shown in Table I. However, the $\Delta E_{\mathrm{r}}(\tau_1)/E_0$ of the lightly doped graphene in Fig. 2(d) changes significantly by the presence of the low charged impurity concentration $n_{\mathrm{i}}=0.17 \times 10^{12}\,\mathrm{cm^{-2}}$, indicating a crossover from the negative $\Delta E_{\mathrm{r}}(\tau_1)/E_0$  to the positive one and the reduction of the carrier scattering due to the enhanced carrier screening effect.
 	Therefore, the information of the accurate charged impurity concentration is required to derive the $\left\langle D_{\textbf{K}}^{2}\right\rangle_{\mathrm{F}}$ from $\Delta E_{\mathrm{r}}(\tau_1)/E_0$ of lightly doped graphene. These findings indicate that heavily doped graphene is suitable for the determination of $\left\langle D_{\textbf{K}}^{2}\right\rangle_{\mathrm{F}}$ from $\Delta E_{\mathrm{r}}(\tau_1)/E_0$.

	\section{Experimental results}
	The graphene sample (Graphene Platform Corporation) that was examined in this study was prepared using chemical vapor deposition. The single-layer graphene (area: 10 mm$\times$10 mm) was transferred to a PET substrate. Raman scattering measurements confirmed the single-layer thickness of the sample and their low defect density. 
	\begin{figure}[b]
		\centering
		\includegraphics[width=8.5cm, bb=0	0	322	129] {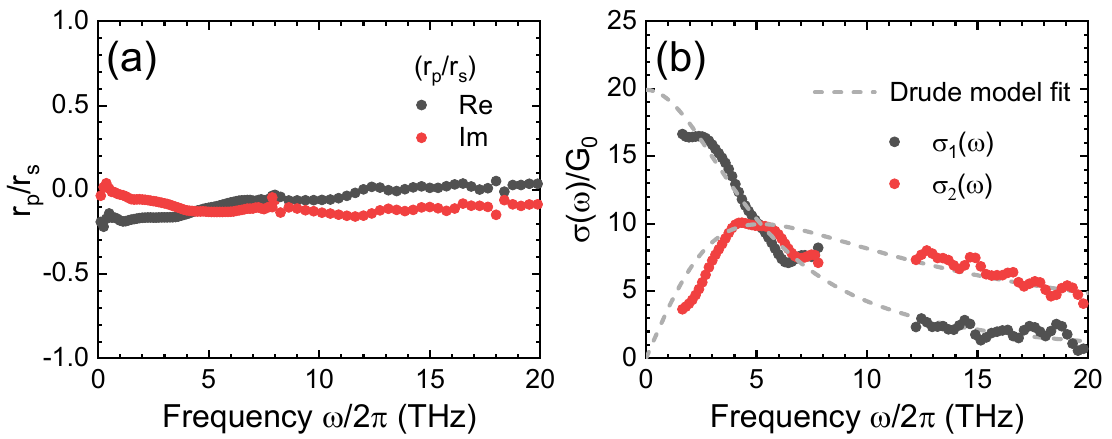}
		\caption{ (a) $(r_{\mathrm{p}}/r_{\mathrm{s}})$ of graphene at equilibrium measured by THz-TDSE. (b) $\sigma(\omega)$ at equilibrium. The dotted line is the fitting curve of the simple Drude model.}
		\label{fgr:example}
	\end{figure}
	The equilibrium THz conductivity of the sample at room temperature ($T_0=295\,{\mathrm{K}}$) was characterized by ultrabroadband THz time domain spectroscopic ellipsometry (THz-TDSE) (see Section I in the SM for details), which enabled the broad Drude peak to be captured directly by measuring the ratio of the reflection coefficient $r_{\mathrm{p}}(\omega)/r_{\mathrm{s}}(\omega)$ in the frequency range between 1.0 and 20 THz
	\cite{Yamashita2014a}, as illustrated in Fig. 4. 
	The fitting of the THz conductivity spectrum obtained from $r_{\mathrm{p}}(\omega)/r_{\mathrm{s}}(\omega)$ by the Drude model allows us to determine the Drude weight $D_0$ and carrier relaxation rate $\Gamma_0$ for the equilibrium state at room temperature $T_0=295\,\mathrm{K}$ accurately. We estimated $D_0=1.36 \times 10^3 G_0$ and $\Gamma_0=21.4\,\mathrm{meV}$, respectively. Here, $G_0=2 e^2/h$ is the quantum conductance. The corresponding Fermi energy is $|\varepsilon_{\mathrm{F}}|=0.43\,\mathrm{eV}$, indicating that the sample is heavily doped and suitable for estimating the EPC strength. The carrier concentration $n_{\mathrm{c}}$ at $T_e=0\,\mathrm{K}$ and the DC conductivity at $T_0$ were estimated as $n_\mathrm{c}=1.1 \times 10^{13} \mathrm{cm^{-2}}$ and $\sigma_{\mathrm{DC}}=20G_0$, respectively, where we used $v_{\mathrm{F}}=1.1 \times 10^6\,\mathrm{m\,s^{-1}}$ considering the carrier and dielectric screening effect in heavily doped graphene on PET substrate\cite{Elias2011a}. 
	
	\begin{figure} [b]
	\centering
	\includegraphics[width=8.6cm, bb=0 0 342 267] {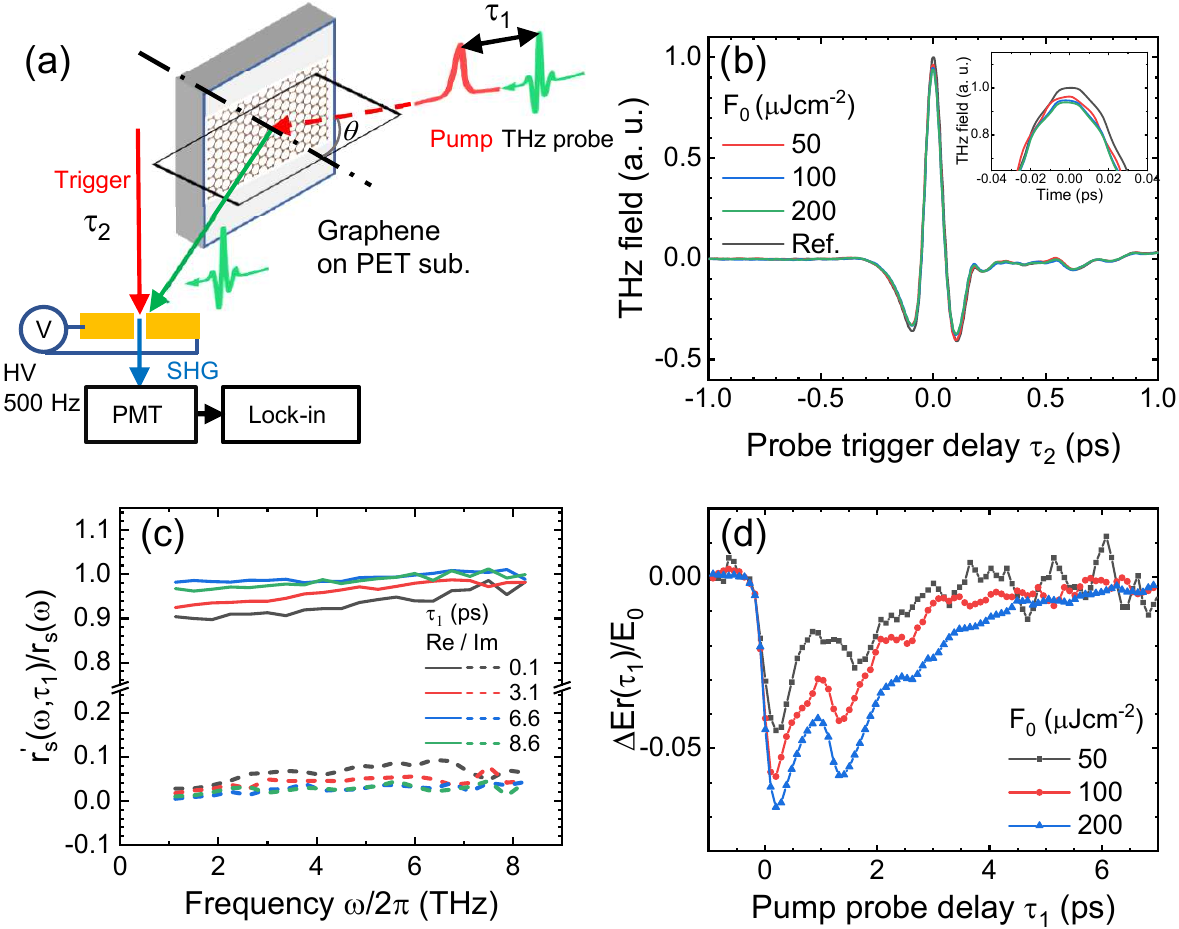}
	\caption{ (a) Schematic of reflection-type OPTP setup. Pump: pump pulse, Trigger: trigger pulse, SHG: second harmonic generation, HV: high voltage, PMT: photomultiplier tube. (b) Temporal waveforms of THz probe pulse measured at $\tau_1=$ 0.1 ps. (c) Frequency dependence of $(r_{\mathrm{s}}^{\prime}(\omega, \tau_1)/r_{\mathrm{s}}(\omega))$ at $\tau_1$ = 0.1, 3.1, 6.6, and 8.6 ps at $F_0=200\,\mu J\,cm^{-2}$. (d) Pump fluence dependence of $\Delta E(\tau_1)/E_0$.}
	\label{fgr:example}
	\end{figure}
	Figure 5(a) presents the optical setup of the reflection-type OPTP used in the experiment. Amplified femtosecond laser pulses (1kHz repetition rate, 785 nm center wavelength) are used to generate ultrabroadband THz probe pulses from laser-excited air plasma\cite{Xie2006}.  S-polarized pump pulses with a pulse duration of 220 fs are loosely focused and excited the graphene sample at an incident angle of $\theta=60^{\circ} $ and the created hot carrier state was probed by s-polarized THz pulses with a pump probe time delay $\tau_1$. The temporal waveforms of the reflected THz probe pulses are measured by air breakdown coherent detection, which detects the second harmonic generation of the trigger pulse induced by the THz electric field\cite{Dai2006}. Figure 5(b) depicts the temporal waveforms of the THz probe pulse reflected from the photoexcited graphene. When the pump fluence is increased, the peak amplitude of THz probe decreases slightly, indicating negative photoconductivity. The ratio of the reflection coefficient $r_{\mathrm{s}}^{\prime}(\omega, \tau_1)/r_{\mathrm{s}}(\omega)$ of graphene with and without pump fluence $F_0=200 \,\mathrm{\mu J\,cm^{-2}}$ calculated by Fourier transformation of the THz waveforms at different $\tau_1$ values, as plotted in Fig. 5(c), decreases and then recovers to the equilibrium reflecting the rise  and subsequent relaxation process of the hot carrier dynamics, and this was used for the calculation of $\sigma(\omega, \tau_1)$ (see Section II in the SM for details).
	Figure 5(d) presents the fluence dependence of $\Delta E_{\mathrm{r}}(\tau_1)/E_0$, which exhibits multiple negative peaks around $\tau_1=0.2,1.4, 2.3\,\mathrm{ps}$ owing to the multiple reflections inside the PET substrate. As $F_0$ increases, the peak height $\Delta E_{\mathrm{r}}(\tau_1)/E_0$ increases but it exhibits saturation behavior with an increased relaxation time. 
	\begin{figure} 
		\centering
		\includegraphics[width=8cm, bb=0 0 316 287] {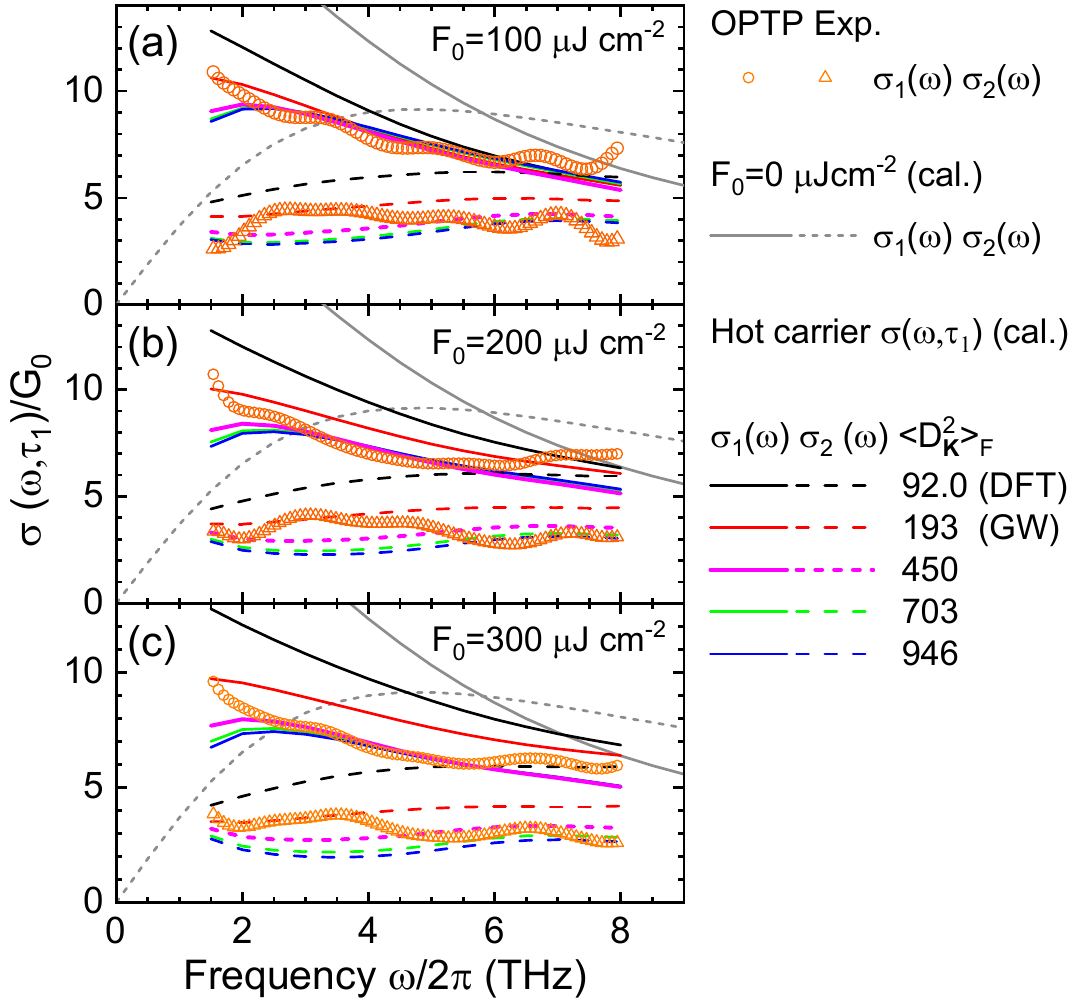}
		\caption{ Pump fluence dependence of $\sigma(\omega, \tau_1)$ (orange symbols) of heavily doped graphene obtained from for $F_0=$(a)100, (b)200 and (c)300$\,\mathrm{\mu J \,cm^{-2}}$. The gray solid and dashed lines correspond to the calculated $\sigma(\omega)$ at equilibrium. 
		The black, red, purple, green and blue lines correspond to $\sigma(\omega, \tau_1)$ at $\tau_1=0.1$ ps using $\left\langle D_{\textbf{K}}^{2}\right\rangle_{\mathrm{F}}=92.0, 193, 450, 703$ and  946\,$(\mathrm{eV\,\AA^{-1})^2}$, respectively.}
		\label{fgr:example}
	\end{figure}

	 Figures\,6(a)-(c) present the pump fluence dependence of $\sigma (\omega, \tau_1)$ measured at $\tau_1=0.1\,\mathrm{ps}$. We observe the reduction of the THz conductivity indicating the large negative photoconductivity with non-Drude behavior as $F_0$ increases and $\sigma (\omega, \tau_1)$ for $F_0=200\,\mathrm{\mu J\,cm^{-2}}$ reaches less than half of that at the equilibrium (gray curve), indicating a significant increase in the carrier scattering  by SCOPs at high temperatures. 
	It is found that $\sigma (\omega, \tau_1)$ for $\left\langle D_{\mathbf{\Gamma}}^{2}\right\rangle_{\mathrm{F}}$ by the DFT (black curve) and GW (blue curve) calculations can not reproduce the observed negative photoconductivity, even if the SA effect is not considered. On the other hand, $\sigma (\omega, \tau_1)$ for $\left\langle D_{\mathbf{\Gamma}}^{2}\right\rangle_{\mathrm{F}}=703$ and $946\,(\mathrm{eV\,\AA^{-1})^2}$ show the larger deviation than that for $\left\langle D_{\mathbf{\Gamma}}^{2}\right\rangle_{\mathrm{F}}=450\,(\mathrm{eV\,\AA^{-1})^2}$. 
	
	Figures 7(a)-(c) depict the comparison of $\Delta E_{\mathrm{r}} (\tau_1)/E_0$ between the experiment and calculations, which is significantly dependent on $\left\langle D_{\mathbf{K}}^{2}\right\rangle_{\mathrm{F}}$ and the pump fluence $F_0$. 
	For $\left\langle D_{\mathbf{K}}^{2}\right\rangle_{\mathrm{F}}$ by the DFT and GW calculations, the peak height and temporal evolution of $\Delta E_{\mathrm{r}}(\tau_1)/E_0$ differ significantly from the experimental values and the higher values $\left\langle D_{\mathbf{K}}^{2}\right\rangle_{\mathrm{F}}=450$--$946 \,(\mathrm{eV\,\AA^{-1}})^2$ are required to reproduce the $\Delta E_{\mathrm{r}}(\tau_1)/E_0$.
	By comparing $\sigma (\omega, \tau_1)$ and $\Delta E_{\mathrm{r}}(\tau_1)/E_0$ with the calculation in Figs.\,6 and 7, we estimated $\left\langle D_{\mathbf{K}}^{2}\right\rangle_{\mathrm{F}}  \approx 450 \,(\mathrm{eV\,\AA^{-1}})^2$ and $\tau_{\mathrm{ie}}=116\,\mathrm{fs}$, 	at which the calculation (blue curves) best fits the experimental results. 
	\begin{figure} [t]
		\centering
		\includegraphics[width=6.5cm, bb=0 0 360 424] {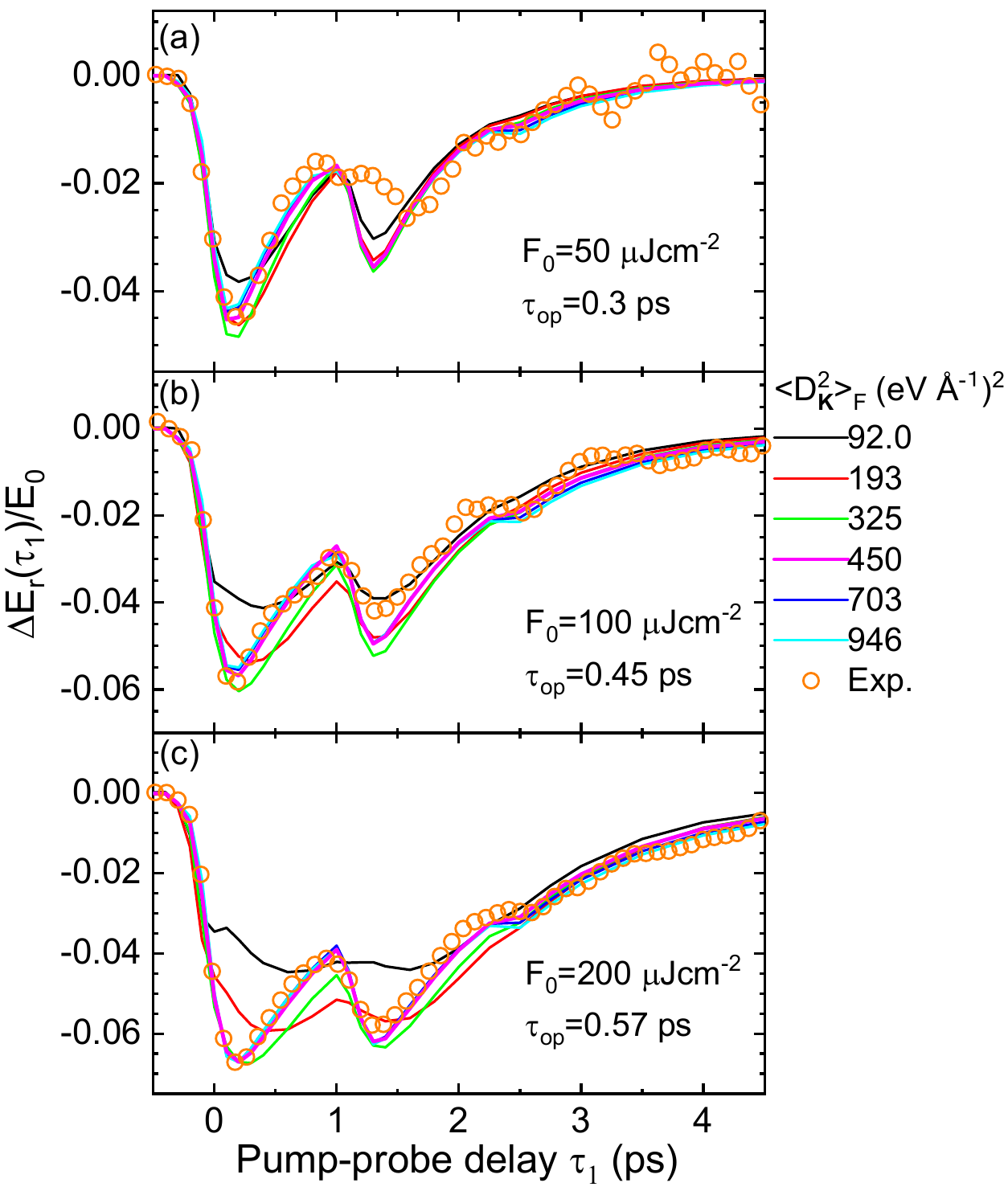}
		\caption{Comparison of $\Delta E_{\mathrm{r}}(\tau_1)/E_0$ between experiment and calculations for different EPCs for pump fluence $F_0=$(a)50, (b)100 and (c)200$\,\mathrm{\mu Jcm^{-2}}$. The red open circles represent the experimental $\Delta E_{\mathrm{r}}(\tau_1)/E_0$. The solid curves correspond to the $\Delta E_{\mathrm{r}}(\tau_1)/E_0$ 
		calculated using $\left\langle D_{\mathbf{\Gamma}}^{2}\right\rangle_{\mathrm{F}}=92.0 \mathrm{\,(DFT, black)}$, $193\,(\mathrm{GW, red})$, $\mathrm{325\,(green)}$, $\mathrm{450\,(purple)}$, $\mathrm{703\,(blue)}$ and $\mathrm{946\,(light\,blue)}\,(\mathrm{eV\,\AA^{-1})^2}$, respectively.}
		\label{fgr:example}
	\end{figure}	
	\begin{figure} 
		\centering
		\includegraphics[width=6.5cm, bb=0 0 208 245] {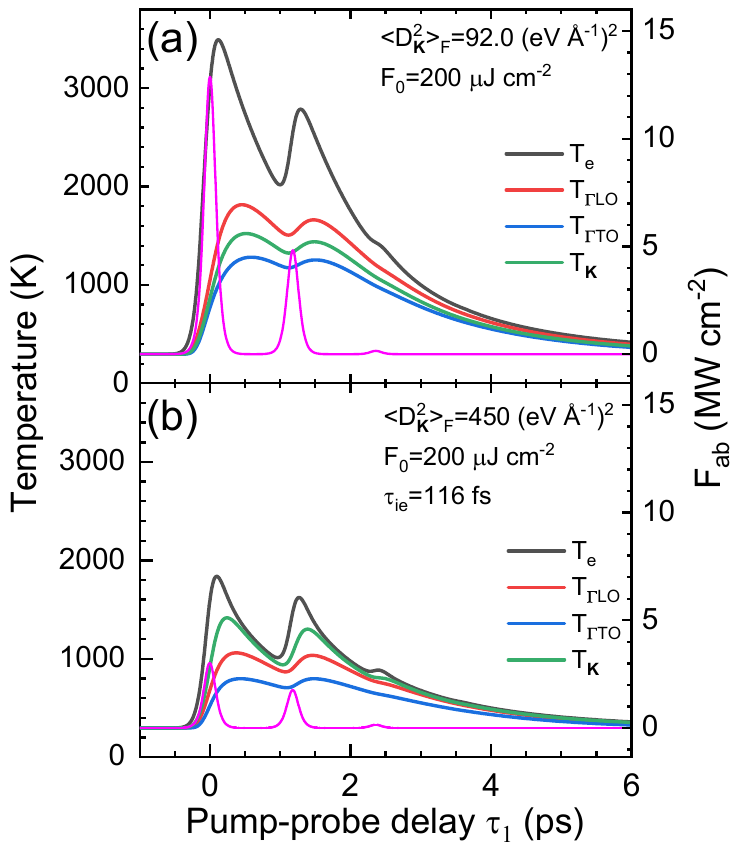}
		\caption{ Temporal evolution of $T_{e}$ and $T_{\eta}$ calculated for $\left\langle D_{\mathbf{K}}^{2}\right\rangle_{\mathrm{F}}=$ (a) 92.0and (b) $450\,(\mathrm{eV\,\AA^{-1})^2}$. The red curve is the absorbed pump intensity $F_{\mathrm{ab}}$, calculated considering the SA effect.}
		\label{fgr:example}
	\end{figure}	
	In this case, the obtained $\tau_{\mathrm{ie}}=116\,\mathrm{fs}$ corresponds to the saturated pump intensity $I_{s}=1.0$ and $1.7 \times 10^8\,\mathrm{W\,cm^{-2}}$ for 
	$\alpha_{\mathrm{inter}}$ and $A^{\mathrm{s*}}_{12}$ respectively, which is slightly smaller than the reported value in Ref.\,\cite{Hyung2012, Marini2017a}.
	
	Figure\,8 presents the temporal evolution of $T_{e}$ and $T_{\eta}$ calculated for $\left\langle D_{\mathbf{K}}^{2}\right\rangle_{\mathrm{F}}=92.0$ and $450\,(\mathrm{eV\,\AA^{-1})^2}$ under the pump fluence 
	$F_0=200\,\mathrm{\mu J\,cm^{-2}}$ indicating that hot carrier and phonon dynamics are significantly dependent on the EPC.  For $\left\langle D_{\mathbf{K}}^{2}\right\rangle_{\mathrm{F}}=92.0\,(\mathrm{eV\,\AA^{-1})^2}$ as shown in Fig.\,8(a), the hot carrier temperature increases beyond $T_e$=3000\,K, and $T_{\mathbf{K}}$ followed $T_e$ slowly owing to the weak EPC and reaches up to $T_{\mathbf{K}} \approx 1500\,\mathrm{K}$. In this high temperature range, the carrier scattering by optical phonons is dominant and the Drude weight $D(T_e)$ makes the positive contribution to $\sigma(\omega, \tau_1)$ in contrast to the carrier scattering. The competition of these factors leads to broader peaks of $\Delta E_{\mathrm{r}}(\tau_1)/E_0$ for DFT (black line) in Fig. 7(c) than those of $T_{\mathrm{\eta}}$ in Fig.\,8(a). For $\left\langle D_{\mathbf{K}}^{2}\right\rangle_{\mathrm{F}}=450\,(\mathrm{eV} \mbox{\AA}^{-1})^2$ as seen in Fig.\,8(b), the hot carrier temperature increases up to only $T_e \approx $2000\,K and $T_{\mathbf{K}}$ follows $ T_e$ rapidly and reaches up to $T_{\mathbf{K}} \approx 1400\,K$ owing to the SA effect and strong EPC. In this case, $D(T_e)$ makes the same contribution to $\sigma(\omega, \tau_1)$ as the optical phonon scattering, resulting in sharper peaks of $\Delta E_{\mathrm{r}}(\tau_1)/E_0$ and a successful reproduction of the experimental results.
	Furthermore, the frequency dependence of $\sigma (\omega, \tau_1)$ at $\tau_1$ = 0.1\,ps in Fig.\,6 deviates from the simple Drude model as $F_0$ increases. This originates from the rapid temporal variation in the carrier temperature and scattering rate during the THz probing time following the photoexcitation, and the calculation with $\left\langle D_{\mathbf{K}}^{2}\right\rangle_{\mathrm{F}}=450 \,(\mathrm{eV\AA^{-1}})^2$ effectively reproduces the observed large negative photoconductivity with non-Drude behavior. This indicates that most photoexcited carriers are recombined and the quasi-equilibrium hot carrier state is almost established at $\tau_1=0.1\,\mathrm{ps}$ owing to the strong Auger recombination in the heavily doped graphene, as reported in Ref. \cite{Gierz2013c}. The parameters used in the calculation are displayed in Table II. 
		
	\begin{table}[b]
	\caption{Parameters used in calculation of $\sigma (\omega, \tau_1)$ and $\Delta E_{\mathrm{r}}(\tau_1)/E_0$ in Figs. 6, 7 and 8\textsuperscript{\emph{a}}.}
	\label{table:data_type}
	\centering
	\begin{tabular}{cccccccc}
		\hline
		$\left\langle D_{\mathrm{K}}^{2}\right\rangle_{\mathrm{F}}$ $(\mathrm{eV} \mbox{\AA}^{-1})^2 $  & $\tau_{\mathrm{ie}}$ (fs)  & $I_{\mathrm{s}} (\mathrm{W\,cm^{-2}})$ &  $n_{\mathrm{i}}\,(\mathrm{cm^{-2}})$  & $\lambda_{\mathbf{K}} (\varepsilon_{\mathrm{F}})$ &  \\
		\hline \hline
		92.0 &  --  & --   & $1.15\times10^{12}$   & 0.02 \\
		193  &  --  & --    & $1.13\times10^{12}$ & 0.04 \\
		450 &  116 & $1.72 \times 10^8$   & $1.09\times10^{12}$  & 0.09\\
		703 &  210 & $0.53\times 10^8$ & $1.05\times10^{12}$  & 0.14\\
		946 &  299 & $0.26\times 10^8$ & $1.01\times10^{12}$ & 0.19\\
		\hline
	\end{tabular}\\
	\textsuperscript{\emph{a}}  The values of $\left\langle D_{\bf{\Gamma}}^{2}\right\rangle_{\mathrm{F}}$, $J_{\mathrm{a}}$ and $\rho_{\mathrm{s}}$ are set to $\left\langle D_{\mathbf{\Gamma}}^{2}\right\rangle_{\mathrm{F}}=45.6\,(\mathrm{eV} \mbox{\AA}^{-1})^2$, $J_{\mathrm{a}}=30.0\,(\mathrm{eV})$ and $\rho_{\mathrm{s}}=40.0\,(\Omega) $, respectively. The charged impurity concentration $n_{\mathrm{i}}$ is selected to provide the same DC conductivity $\sigma_{\mathrm{DC}}=20.0 G_0$ at equilibrium for $T_0=295\,\mathrm{K}$. 
	\end{table}

	\section{Discussion}
	 Based on the fitting of $\Delta E_{\mathrm{r}}(\tau_1)/E_0$ by the calculation considering the EPC, we estimated the phenomenological phonon decay time due to lattice anharmonicity as $\tau_{\mathrm{ph}}=0.3, 0.45$ and $0.57\,\mathrm{ps}$ for $F_0=50, 100$ and 
	 $200\,\mathrm{\mu J\,cm^{-2}}$, respectively. Refs. \cite{Kang2010, Gao2011} reported longer $\tau_{\mathrm{ph}}= 0.8$--1.5$\,\mathrm{ps}$ for graphene on a $\mathrm{SiO_2}$ substrate. However, these values were determined from the simple fitting of transient absorption or anti-stokes Raman intensity by exponential function and do not consider the EPC. The simple fitting of $\Delta E_{\mathrm{r}}(\tau_1)/E_0$ with exponential curve results
	 in $\tau_{\mathrm{ph}}=$ 1.15--1.5\,ps which are comparable to the reported values. The theoretical study reported the phonon decay time $\tau_{\mathrm{ph}} \approx$ 3.5 and 4.5\,ps for $\mathbf{\Gamma}$ and $\mathbf{K}$ phonon by only considering the anharmonicity of lattice in graphene without substrate \cite{Bonini2007}. Therefore, the obtained $\tau_{\mathrm{ph}}$  indicates the dominant contribution of substrate for the optical phonon decay channel.

	\begin{figure} [t]
	\centering
	\includegraphics[width=6.5cm, bb=0 0 215 170] {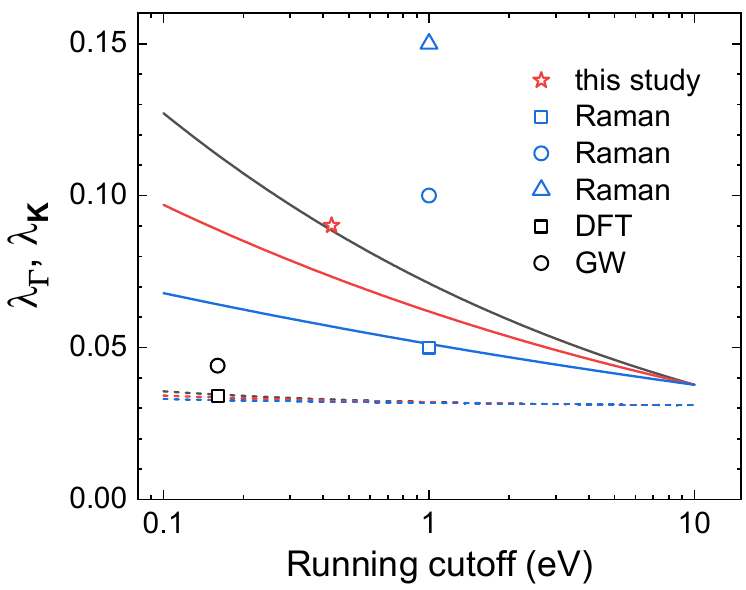}
	\caption{ Flow of dimensionless coupling constants $\lambda_{\mathbf{\Gamma}}$ and $\lambda_{\mathbf{K}}$ (three dashed and solid curves, respectively) for three values of $\epsilon_{\mathrm{av}}$=1, 2, and 5. The red and blue symbols correspond to the $\lambda_{\mathbf{K}}$ determined 
	in this study ($\epsilon_{\mathrm{av}}=2$) and by Raman studies ($\epsilon_{\mathrm{av}}=5$) from Ref. \cite{Basko2009a} (blue open square and circle) and Ref. \cite{Froehlicher2015} (blue open triangle). The black open circle and square correspond to $\lambda_{\mathbf{K}}$ by the DFT \cite{Piscanec2004a, Basko2009a} and GW \cite{Lazzeri2008, Gruneis2008, Basko2009a} calculations ($\epsilon_{\mathrm{av}}=1$), respectively.}
	\label{fgr:example}
	\end{figure}	

	The dimensionless coupling constants $\lambda_{\mathbf{\Gamma}}$ and $\lambda_{\mathbf{K}}$ for the optical phonons near the $\mathbf{\Gamma}$ and $\mathbf{K}$ points, respectively, are useful for comparing the EPC strengths determined from various experiments and calculations, which are defined as \cite{Basko2009a}
	\begin{equation}
		\lambda_{\mathbf{\Gamma},\mathbf{K}}=\frac{F^2_{\mathbf{\Gamma},\mathbf{K}} A_{\mathrm{u.c.}}}{2M \hbar \omega_{\mathbf{\Gamma},\mathbf{K}}  v_F^2}.
	\end{equation}
	In the above, $M \approx2.00 \times 10^{-26} \,\mathrm{kg}$ is the mass of the carbon atom and $A_{\mathrm{u.c.}} \approx 5.24 \,\mbox{\AA}^2$ is the unit-cell area. $F^2_{\mathbf{\Gamma}} $ and $F^2_{\mathbf{K}} $ have the dimensionality of a force and are the proportionality coefficients between the change in the effective Hamiltonian and lattice displacement along the corresponding phonon mode. Subsequently, 
	the matching rules are expresses as $F^2_{\mathbf{\Gamma}}=4\left\langle D_{\mathbf{\Gamma}}^{2}\right\rangle_{\mathrm{F}}$ and $F^2_{\mathbf{K}}=2\left\langle D_{\mathbf{K}}^{2}\right\rangle_{\mathrm{F}}$.
	Note that $\lambda_{\mathbf{K}}$ is subject to Coulomb renormalization, which implies that $\lambda_{\mathbf{K}}$ is dependent on the electronic energy scale, such as the electron energy, Fermi energy, or temperature T, whichever is larger: $\lambda_{\mathbf{K}}=\lambda_{\mathbf{K}}(\mathrm{max} \{|\varepsilon|, |\varepsilon_{\mathrm{F}}|,|T|\})$.  
	From $\left\langle D_{\mathbf{K}}^{2}\right\rangle_{\mathrm{F}}  \approx 450 \,\mathrm{(eV\,A^{-1})^2}$, we estimated $\lambda_{\mathbf{K}} (\varepsilon_{\mathrm{F}}) \approx 0.09$ using Eqs. (13) and (14). 
	Figure 9 presents the flow of $\lambda_{\mathbf{\Gamma}}$ and $\lambda_{\mathbf{K}}$ for different background static dielectric 
	constants $\epsilon_{\mathrm{av}}=(1+\epsilon_{\mathrm{s}})/2=1$, 2, and 5 calculated by solving the renormalization group equation in Ref. \cite{Basko2008e}, which sum up the leading logarithmic corrections and go beyond the Hartree--Fock approximation. 
	The bare values of the dimensionless EPCs $\lambda_{\mathbf{\Gamma}}=0.031$ and $\lambda_{\mathbf{K}}=0.038$ were selected to satisfy the relation $\lambda_{\mathbf{\Gamma}}/\lambda_{\mathbf{K}}=\omega_{\mathbf{K}}/\omega_{\mathbf{\Gamma}}$ and to reproduce the experimental value  $\lambda_{\mathbf{\Gamma}}=0.031$ \cite{Froehlicher2015}. 
	The renormalization group analysis demonstrated that, although $\lambda_{\mathbf{\Gamma}}$ was almost constant, $\lambda_{\mathbf{K}}$ was strongly dependent on the energy scale as well as $\epsilon_{\mathrm{av}}$.
	The obtained $\lambda_{\mathbf{K}} (\varepsilon_{\mathrm{F}})=0.09$ slightly larger than the calculated value of $\lambda_{\mathbf{K}}(\varepsilon_{\mathrm{F}})=0.073$. According to the ratio of the $\lambda ({\omega_\mathbf{K}})/\lambda_{\mathbf{K}}(\varepsilon_{\mathrm{F}})=1.21$ for $\epsilon_{\mathrm{av}}=2$ in Fig. 9, we obtained $\lambda ({\omega_\mathbf{K}})=0.11$, which is a factor of 3.2 larger than the DFT value $\lambda_{\mathbf{K}}({\omega_\mathbf{K}})=0.034$. Raman studies \cite{Ferrari2006, Das2008b, Das2009, Berciaud2009, Froehlicher2015} using a field effect transistor based on the polymer electrolyte ($\epsilon_{\mathrm{av}}=5$) reported $\lambda_{\mathbf{\Gamma}}=$0.028 and 0.031 from the ratio of the area between G and the 2D peak, which were comparable to $\lambda_{\mathbf{\Gamma}}=0.028$ by the DFT calculation of $\left\langle g_{\mathbf{\Gamma}}^{2}\right\rangle_{\mathrm{F}}$ using Eq. (14). However, $\lambda_{\mathbf{K}}(E_{\mathrm{L}}/2)$ ranged between 0.05 and 0.15 as seen in Fig. 9, where $E_L$ is the laser excitation energy (for a typical Raman measurement $E_{\mathrm{L}}/2 \sim 1 \mathrm{eV}$). The corresponding $\lambda_{\mathbf{K}}(\omega_{\mathbf{K}})$ are estimated as 0.063 and 0.19. The lower limit value is comparable to the calculated $\lambda_{\mathbf{K}}(\omega_{\mathbf{K}})$ for $\epsilon_{\mathrm{av}}=5$. Although Raman spectroscopy is a powerful tool for the determination of $\lambda_{\mathbf{K}}(\omega_{\mathbf{K}})$ as well as $\lambda_{\mathbf{\Gamma}}(\omega_{\mathbf{K}})$, it requires the accurate estimation of the gate capacitance of FET device which are not required in OPTP experiments. 
	
	\section{Conclusion}
	In conclusion, we investigated the EPC of the optical phonons near the $\mathbf{K}$ point of heavily doped graphene on PET substrate and the hot carrier dynamics using a combination of the time-resolved THz spectroscopy and numerical simulations. The hot carrier dynamics in heavily doped graphene on PET substrate is less sensitive to the extrinsic charged impurity and surface polar phonons of the substrate and is dominated by the electron-optical phonon interactions. 
	According to the quantitative analysis based on the BTE and comprehensive temperature model considering the SA effect on pump fluence, the $\Delta E( \tau_1)/E_0$ value can be used for the determination of the EPC in graphene. The estimated  $\left\langle D_{\mathbf{K}}^{2}\right\rangle_{\mathrm{F}}  \approx 450 \,\mathrm{(eV \AA^{-1})^2}$ indicates the strong renormalization by e-e interaction and the corresponding dimensionless coupling constant $\lambda_{\mathbf{K}}(E_\mathrm{F}) \approx 0.09$ slightly larger than the calculation by the renormalization group theory. The extension of  the simulation model for the 
	undoped or lightly doped graphene on various substrate requires the accurate estimation of charged impurities and surface polar phonons of the substrate is a future issue that will be important to the development of graphene optoelectronic devices. 

\begin{acknowledgments}
		This work was supported by the JSPS KAKENHI (19H01905) and Research Foundation for Opto-Science and Technology.
	
\end{acknowledgments}
	
	
	
\bibliographystyle{apsrev4-1.bst}
\bibliography{library}

\end{document}


\title{Supplemental material\\ 
			Hot carriers dynamics and carrier-phonon interactions in graphene studied by ultra-broadband time resolved THz spectroscopy}
	
	\author{Sho Ikeda}%
	\author{Chiko Otani}%
	\author{Masatsugu Yamashita}%
	\email[e-mail: ]{m-yama@riken.jp}
	\affiliation{Terahertz Sensing and Imaging Team, RIKEN Center for Advanced Photonics, 519-1399 Aramaki-Aoba Aoba-ku, Sendai, Miyagi 980-0845, Japan}
	\date{26 March 2021}%
		
	\maketitle
	\newpage

\section{Calculation of equilibrium optical conductivity of graphene from THz time domain spectroscopic ellipsometry experiment} 
In this section, we explain the calculation procedure of the THz conductivity $\sigma (\omega_{\mathrm{THz}})$ of graphene on the substrate from the ratio of the complex reflection coefficient $(r_{\mathrm{p}}(\omega_{\mathrm{THz}})/r_{\mathrm{s}}(\omega_{\mathrm{THz}}))$ for the p- and s- polarized THz waves measured by THz time domain spectroscopic ellipsometry (THz-TDSE)\cite{Yamashita2014a}. 
According to the standard thin-film approximation, the reflection coefficients of graphene on a substrate for p- and s-polarized THz wave are given by\cite{Bludov2013a}
\begin{subequations}
\begin{equation}
	\begin{aligned}
		r_{\mathrm{p}}(\omega_{\mathrm{THz}}) =\frac{\sigma (\omega_{\mathrm{THz}}) Z_0+(\frac{\epsilon_{\mathrm{2}}(\omega_{\mathrm{THz}})}{(\epsilon_{\mathrm{2}}(\omega_{\mathrm{THz}})-\epsilon_{\mathrm{1}}(\omega_{\mathrm{THz}}) \sin^2 \theta_1)^{1 / 2}}-\frac{\epsilon^{1/2}_{\mathrm{1}}(\omega_{\mathrm{THz}}))}{\cos \theta_1})}{\sigma (\omega_{\mathrm{THz}}) Z_0+(\frac{\epsilon_{\mathrm{2}}(\omega_{\mathrm{THz}})}{(\epsilon_{\mathrm{2}}(\omega_{\mathrm{THz}})-\epsilon_{\mathrm{1}}(\omega_{\mathrm{THz}}) \sin^2 \theta_1)^{1 / 2}}+\frac{\epsilon^{1/2}_{\mathrm{1}}(\omega_{\mathrm{THz}})}{\cos \theta_1})},
	\end{aligned}
\end{equation}
\begin{equation}
	\begin{aligned}
		r_{\mathrm{s}}(\omega_{\mathrm{THz}})  =-\frac{\sigma (\omega_{\mathrm{THz}}) Z_0+(\epsilon_{\mathrm{2}}(\omega_{\mathrm{THz}})-\epsilon_{\mathrm{1}}(\omega_{\mathrm{THz}}) \sin^2 \theta_1)^{1 / 2}-\epsilon_{1}^{1 / 2}(\omega_{\mathrm{THz}}) \cos \theta_1}{\sigma (\omega_{\mathrm{THz}}) Z_0+(\epsilon_{\mathrm{2}}(\omega_{\mathrm{THz}})-\epsilon_{\mathrm{1}}(\omega_{\mathrm{THz}}) \sin^2 \theta_1)^{1 / 2}+\epsilon_{1}^{1 / 2}(\omega_{\mathrm{THz}}) \cos \theta_1}.
	\end{aligned}
\end{equation}
\end{subequations}
In the above, $Z_0=376.7\,(\Omega)$ is the vacuum impedance and $\theta_1=60^{\circ}$ is the incidence angle of the THz wave. Furthermore, $\epsilon_{\mathrm{i}}(\omega_{\mathrm{THz}})$ is the dielectric constant of layer i, as indicated in Fig.S1. From Eq. (SI.1), $\sigma (\omega)$ is expressed as
\begin{equation}
		\sigma (\omega) = -\frac{\{(r_{\mathrm{p}}/r_{\mathrm{s}})(A+B^{\prime})+A^{\prime}+B\}Z_0+\{((r_{\mathrm{p}}/r_{\mathrm{s}})(A+B^{\prime})+A^{\prime}+B)^2-4(1+(r_{\mathrm{p}}/r_{\mathrm{s}}))((r_{\mathrm{p}}/r_{\mathrm{s}}) A B^{\prime}+A^{\prime} B)\}^{1/2}}{2(1+(r_{\mathrm{p}}/r_{\mathrm{s}}))Z_0},
\end{equation}
where
\begin{subequations}
\begin{equation}
	A=\frac{\epsilon_{\mathrm{2}}(\omega_{\mathrm{THz}})}{(\epsilon_{\mathrm{2}}(\omega_{\mathrm{THz}})-\epsilon_{\mathrm{1}}(\omega_{\mathrm{THz}}) \sin^2 \theta_1)^{1 / 2}}+\frac{\epsilon^{1/2}_{\mathrm{1}}(\omega_{\mathrm{THz}})}{\cos \theta_1},
\end{equation}
\begin{equation}
	A^{\prime}=\frac{\epsilon_{\mathrm{2}}(\omega_{\mathrm{THz}})}{(\epsilon_{\mathrm{2}}(\omega_{\mathrm{THz}})-\epsilon_{\mathrm{1}}(\omega_{\mathrm{THz}}) \sin^2 \theta_1)^{1 / 2}}-\frac{\epsilon^{1/2}_{\mathrm{1}}(\omega_{\mathrm{THz}})}{\cos \theta_1},
\end{equation}
\begin{equation}
	B=(\epsilon_{\mathrm{2}}(\omega_{\mathrm{THz}})-\epsilon_{\mathrm{1}}(\omega_{\mathrm{THz}}) \sin^2 \theta_1)^{1 / 2}+\epsilon_{1}^{1 / 2}(\omega_{\mathrm{THz}}) \cos \theta_1,
\end{equation}
\begin{equation}
	B^{\prime}=(\epsilon_{\mathrm{2}}(\omega_{\mathrm{THz}})-\epsilon_{\mathrm{1}}(\omega_{\mathrm{THz}}) \sin^2 \theta_1)^{1 / 2}-\epsilon_{\mathrm{1}}^{1 / 2}(\omega_{\mathrm{THz}}) \cos \theta_1.
\end{equation}
\end{subequations}
By substituting $(r_{\mathrm{p}}/r_{\mathrm{s}})$ measured by THz-TDSE into Eq. (SI.2), $\sigma (\omega_{\mathrm{THz}})$ can be determined.
\begin{figure}[h]
	\centering
	\includegraphics[width=10cm, bb=0 0 336 135]{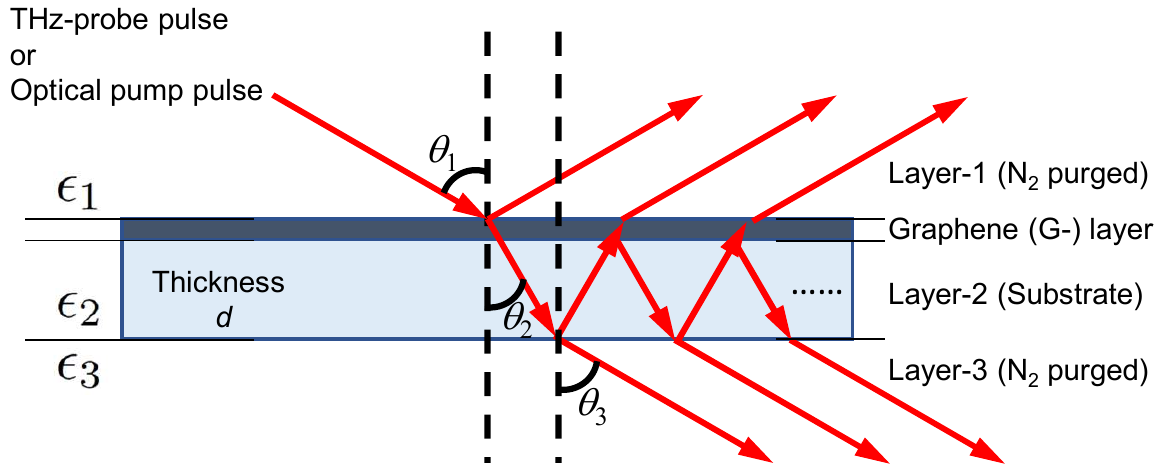}
	\caption{Schematic of systems considered in THz-TDSE and OPTP measurements, showing incident THz probe or optical pump pulse.}
	\label{fgr:example}
\end{figure}

	\newpage
\section{Calculation of hot carrier optical conductivity photoexcited graphene from reflection coefficient by OPTP experiment} 
In this section, we present the calculation procedure of the hot carrier THz conductivity $\sigma(\omega_{\mathrm{THz}}, \tau_1)$ of photoexcited graphene at the pump probe delay $\tau_1$ from the reflection-type OPTP measurement. The reflection-type OPTP measures the ratio of the complex reflection coefficient $X_{\mathrm{s}}(\omega_{\mathrm{THz}},\tau_1)=r_{\mathrm{s}}^{\prime}(\omega_{\mathrm{THz}},\tau_1)/r_{\mathrm{s}}(\omega_{\mathrm{THz}})$ of graphene with and without photoexcitation.
The reflection coefficient for the s-polarization of graphene with complex conductivity $\sigma(\omega_{\mathrm{THz}})$ at an incident angle of $\theta_1$ is expressed by Eq. (SI.1b). Similarly, the THz-amplitude reflection coefficient for the s-polarization of graphene with hot carrier complex conductivity $\sigma(\omega_{\mathrm{THz}}, \tau_1)$ on the substrate at an incident angle of $\theta_1$ for the pump probe delay $\tau_1$ is expressed by
\begin{equation}
	\begin{aligned}
		r_{\mathrm{s}}^{\prime}(\omega_{\mathrm{THz}},\tau_1)=-\frac{\sigma(\omega_{\mathrm{THz}},\tau_1) Z_0+(\epsilon_{\mathrm{2}}(\omega_{\mathrm{THz}})-\epsilon_{\mathrm{1}}(\omega_{\mathrm{THz}}) \sin^2 \theta_1)^{1 / 2}-\epsilon_{1}^{1 / 2}(\omega_{\mathrm{THz}}) \cos \theta_1}{\sigma(\omega_{\mathrm{THz}},\tau_1) Z_0+(\epsilon_{\mathrm{2}}(\omega_{\mathrm{THz}})-\epsilon_{\mathrm{1}}(\omega_{\mathrm{THz}}) \sin^2 \theta_1)^{1 / 2}+\epsilon_{1}^{1 / 2}(\omega_{\mathrm{THz}}) \cos \theta_1}.
	\end{aligned}
\end{equation}
Using Eqs. (SI.1b) and (SII.1), we obtain 
\begin{equation}
	\begin{aligned}
		\sigma(\omega_{\mathrm{THz}},\tau_1) =-\frac{B X_{\mathrm{s}}(\omega_{\mathrm{THz}},\tau_1) r_{\mathrm{s}}(\omega_{\mathrm{THz}})+B^{\prime}}{Z_0[1+ X_{\mathrm{s}}(\omega_{\mathrm{THz}},\tau_1) r_{\mathrm{s}}(\omega_{\mathrm{THz}})]},
	\end{aligned}
\end{equation}
where $B$ and $B^{\prime}$ are provided by Eqs. (SI.3c) and (SI.3d), respectively, and $r_{\mathrm{s}}(\omega_{\mathrm{THz}})$ is calculated using the equilibrium $\sigma (\omega_{\mathrm{THz}})$ obtained by THz-TDSE. We can obtain the $\sigma(\omega_{\mathrm{THz}},\tau_1)$ by substituting $X_{\mathrm{s}}(\omega_{\mathrm{THz}},\tau_1)$ into Eq. (SII.2).

	\newpage
\section{Rate equations for Temperature model } 
In this section, we present the derivation of the hot carrier recombination and generation rate by optical phonon emission and absorption process, respectively, used in the temperature model. The Hamiltonian of electron-phonon interaction $H_{cp}$ is 
\begin{equation} 
	H_{ep}=\sum_{\bm{k}, \bm{k}^{\prime}, \bm{q}} V_{ep}(c^{\dagger}_{\bm{k}}c_{\bm{k}^{\prime}}b_{\bm{q}}+ c^{\dagger}_{\bm{k}^{\prime}}c_{\bm{k}}b_{-\bm{q}}^{\dagger})
\end{equation}
Here, $V_{ep}$ is the potential of the electron-phonon interaction, $c^{\dagger}_{\bm{k}}(c_{\bm{k}})$ is the creation (annihilation ) operator with carrier wave vector $\bm{k}$, $b^{\dagger}_{\bm{q}}(b_{\bm{q}})$ is the creation (annihilation ) operator with phonon wave vector $\bm{q}$.
From Fermi's golden rule, the carrier transition rate from $\bm{k}$ to $\bm{k}^{\prime}$ by the emission and absorption of the $\Gamma_{\mathrm{LO}}$ phonon or $\Gamma_{\mathrm{LO}}$ phonon  with the energy of $\hbar \omega_{\bf{\Gamma}}$ are given by 
\begin{equation} 
	\begin{aligned}
		P^{\mathrm{EM/AB},\bf{\Gamma}}_{\lambda \bm{k} \lambda^{\prime} \bm{k}^{\prime}}&= \frac{2 \pi}{\hbar} \left| \left\langle \bm{k}^{\prime},\lambda^{\prime}\left| H_{ep} \right|\bm{k},\lambda \right\rangle \right| ^2 \delta (\varepsilon_{\lambda^{\prime} \bm{k}^{\prime}}-\varepsilon_{\lambda\bm{k}} \pm \hbar \omega_{\bf{\Gamma}})\\
		&=\frac{\pi\left| \mathrm{D}^{\bf{\Gamma}}_{\lambda \bm{k} \lambda^{\prime} \bm{k}^{\prime} }\right| ^2}{\rho \omega_{\bf{\Gamma}} A} (n_{\bf{\Gamma}}+\frac{1}{2} \pm \frac{1}{2}) \delta (\varepsilon_{\lambda^{\prime} \bm{k}^{\prime}}-\varepsilon_{\lambda\bm{k}} \pm \hbar \omega_{\bf{\Gamma}})  \delta(\bm{k}^{\prime}-\bm{k} \pm \bm{q})\\
	\end{aligned}
\end{equation}
Here, $\left| \mathrm{D}^{\bf{\Gamma}}_{\lambda\bm{k} \lambda^{\prime}\bm{k}^{\prime} }\right|^2$ is the square of the EPC matrix element. For small $\bm{q}$ and $\bm{k}$, the EPC matrix elements are   $\left| \mathrm{D}^{\bf{\Gamma}}_{\lambda \bm{k} \lambda^{\prime}\bm{k}^{\prime} }\right|^2=\left\langle \mathrm{D}_{\bf{\Gamma}}^2 \right\rangle_F \left[ 1 \pm \cos (\theta_{\bm{k},\bm{q}} + \theta_{\bm{k}^{\prime},\bm{q}})\right]$ where $\left\langle \mathrm{D}_{\bf{\Gamma}}^2 \right\rangle_F$ is the average on Fermi surface of $\left| \mathrm{D}^{\bf{\Gamma}}_{\lambda\bm{k}, \lambda^{\prime}\bm{k}^{\prime} }\right|^2$. $\rho$ is the mass density, A is the area of graphene sample, $\varepsilon_{\lambda \bm{k}}=\lambda \hbar v_F |\bm{k}|$ is the energy of 2D MDF and $\lambda=\pm 1$ is the band index. The upper  and lower sign corresponds to the optical phonon emission and absorption process, respectively.
The corresponding hot carrier recombination and generation rate per unit area including both intra- and inter-band transitions are written as
\begin{subequations}
\begin{equation}
	\begin{aligned}
		R_{\bf{\Gamma}}&=\frac{1}{A}\sum_{\lambda, \lambda^{\prime}}\sum_{\bm{k},\bm{k^{\prime}}} P^{\mathrm{EM},\bf{\Gamma}}_{\lambda \bm{k} \lambda^{\prime} \bm{k}^{\prime}} f_{\lambda}(\bm{k})(1-f_{\lambda^{\prime}}(\bm{k}^{\prime}))\\
		&=\frac{1}{A} \sum_{\lambda, \lambda^{\prime}}\sum_{\bm{k}} \sum_{\bm{k}^{\prime}} \frac{\pi \left\langle \mathrm{D}_{\bf{\Gamma}}^2 \right\rangle_F \left[ 1 \pm \cos (\theta_{ \bm{k},\bm{q}}+\theta_{ \bm{k}^{\prime},\bm{q}})\right]  }{\rho\omega_{\bf{\Gamma}} A} (n_{\bf{\Gamma}}+1)f_{\lambda}(\bm{k)} (1-f_{\lambda^{\prime}}(\bm{k}^{\prime}))  \delta (\varepsilon_{\lambda^{\prime} \bm{k}^{\prime}}-\varepsilon_{\lambda\bm{k}}+\hbar \omega_{\bf{\Gamma}})  \delta(\bm{k}^{\prime}-\bm{k}+\bm{q})\\
		&=\sum_{\lambda, \lambda^{\prime}}\sum_{\bm{k}} \frac{\pi \left\langle \mathrm{D}_{\bf{\Gamma}}^2 \right\rangle_F }{A(2\pi)^2}\int d^2\bm{k}^{\prime} \frac{\left[ 1 \pm \cos (\theta_{\bm{k},\bm{q}}+\theta_{\bm{k}^{\prime},\bm{q}}))\right]}{\rho\omega_{\bf{\Gamma}_{\mathrm{LO}}} } (n_{\bf{\Gamma}}+1) f_{\lambda}(\bm{k)} (1-f_{\lambda^{\prime}}(\bm{k}^{\prime}))  \delta (\varepsilon_{\lambda^{\prime} \bm{k}^{\prime}}-\varepsilon_{\lambda\bm{k}}+\hbar \omega_{\bf{\Gamma}})  \delta(\bm{k}^{\prime}-\bm{k}+\bm{q})\\
		&=\sum_{\lambda, \lambda^{\prime}}\frac{\left\langle \mathrm{D}_{\bf{\Gamma}}^2 \right\rangle_F (n_{\bf{\Gamma}}+1)}{4 \pi \rho\omega_{\bf{\Gamma}}}\int f_{\lambda}(\varepsilon_{\lambda \bm{k}}) N(\varepsilon_{\lambda \bm{k}}) d\varepsilon_{\lambda \bm{k}} \\
		&\quad \times \int d^2 \bm{k}^{\prime} \left[ 1 \pm \cos (\theta_{ \bm{k},\bm{q}}+\theta_{\bm{k}^{\prime},\bm{q}})\right]  (1-f_{\lambda^{\prime}}(\bm{k}^{\prime})) \delta (\varepsilon_{\lambda^{\prime} \bm{k}^{\prime}}-\varepsilon_{\lambda\bm{k}}+\hbar \omega_{\bf{\Gamma}})  \delta(\bm{k}^{\prime}-\bm{k}+\bm{q})\\
	\end{aligned}
\end{equation}
\begin{equation}
	\begin{aligned}
		G_{\bf{\Gamma}}&=\frac{1}{A}\sum_{\lambda, \lambda^{\prime}}\sum_{\bm{k},\bm{k^{\prime}}} P^{\mathrm{AB},\bf{\Gamma}}_{\lambda \bm{k} \lambda^{\prime} \bm{k}^{\prime}} f_{\lambda}(\bm{k})(1-f_{\lambda^{\prime}}(\bm{k}^{\prime}))\\
		&=\frac{1}{A} \sum_{\lambda, \lambda^{\prime}}\sum_{\bm{k}} \sum_{\bm{k}^{\prime}} \frac{\pi \left\langle \mathrm{D}_{\bf{\Gamma}}^2 \right\rangle_F \left[ 1 \pm \cos (\theta_{\bm{k},\bm{q}}+\theta_{\bm{k}^{\prime},\bm{q}})\right]  }{\rho\omega_{\bf{\Gamma}} A}n_{\bf{\Gamma}} f_{\lambda}(\bm{k)} (1-f_{\lambda^{\prime}}(\bm{k}^{\prime})) \delta (\varepsilon_{\lambda^{\prime} \bm{k}^{\prime}}-\varepsilon_{\lambda\bm{k}}-\hbar \omega_{\bf{\Gamma}})  \delta(\bm{k}^{\prime}-\bm{k}-\bm{q})\\
		&=\sum_{\lambda, \lambda^{\prime}}\sum_{\bm{k}} \frac{\pi \left\langle \mathrm{D}_{\bf{\Gamma}}^2 \right\rangle_F }{A(2\pi)^2}\int d^2\bm{k}^{\prime} \frac{\left[ 1 \pm \cos (\theta_{ \bm{k},\bm{q}}+\theta_{ \bm{k}^{\prime},\bm{q}}))\right]}{\rho\omega_{\bf{\Gamma}_{\mathrm{LO}}} } n_{\bf{\Gamma}}f_{\lambda}(\bm{k)} (1-f_{\lambda^{\prime}}(\bm{k}^{\prime})) \delta (\varepsilon_{\lambda^{\prime} \bm{k}^{\prime}}-\varepsilon_{\lambda\bm{k}}-\hbar \omega_{\bf{\Gamma}})  \delta(\bm{k}^{\prime}-\bm{k}-\bm{q})\\
		&=\sum_{\lambda, \lambda^{\prime}}\frac{\left\langle \mathrm{D}_{\bf{\Gamma}}^2 \right\rangle_F n_{\bf{\Gamma}}}{4 \pi \rho\omega_{\bf{\Gamma}} }\int f_{\lambda}(\varepsilon_{\lambda \bm{k}}) N(\varepsilon_{\lambda \bm{k}}) d\varepsilon_{\lambda \bm{k}} \\
		&\quad \times \int d^2 \bm{k}^{\prime} \left[ 1 \pm \cos (\theta_{ \bm{k},\bm{q}}+\theta_{\bm{k}^{\prime},\bm{q}})\right]  (1-f_{\lambda^{\prime}}(\bm{k}^{\prime})) \delta (\varepsilon_{\lambda^{\prime} \bm{k}^{\prime}}-\varepsilon_{\lambda\bm{k}}-\hbar \omega_{\bf{\Gamma}})  \delta(\bm{k}^{\prime}-\bm{k}-\bm{q})\\
	\end{aligned}
\end{equation}
\end{subequations}
Here, $N(\varepsilon_{\lambda \bm{k}})=2|\varepsilon_{\lambda \bm{k}}|/\pi (\hbar v_F)^2$ is the density of state of 2D MDF. Furthermore, the electron distribution function $f_{\lambda}(\bm{k})$ can be replaced by Fermi-Dirac type distribution $f_0(\varepsilon_{\lambda \bm{k}}, T_e)$ for hot carriers in quasi-equilibrium. Similarly, the hot carrier recombination and generation rate by K-phonon with the energy of $\hbar \omega_{\bf{K}}$are given by
\begin{subequations}
\begin{equation}
	\begin{aligned}
		R_{\bf{K}_{\mathrm{}}}&=\frac{1}{A}\sum_{\lambda, \lambda^{\prime}}\sum_{\bm{k},\bm{k^{\prime}}} P^{\mathrm{EM},\bf{K}}_{\lambda \bm{k} \lambda^{\prime} \bm{k}^{\prime}} f_{\lambda}(\bm{k})(1-f_{\lambda^{\prime}}(\bm{k}^{\prime}))\\
		&=\sum_{\lambda, \lambda^{\prime}}\frac{\left\langle \mathrm{D}_{\bf{K}}^2 \right\rangle_F (n_{\bf{K}}+1)}{4 \pi \rho\omega_{\bf{K}} }\int f_{\lambda}(\varepsilon_{\lambda \bm{k}}) N(\varepsilon_{\lambda \bm{k}}) d\varepsilon_{\lambda \bm{k}}\\
		&\quad \times \int d^2\bm{k}^{\prime} \left[ 1 \pm \cos (\theta_{\bm{k},\bm{k}^{\prime}})\right]  (1-f_{\lambda^{\prime}}(\bm{k}^{\prime})) \delta (\varepsilon_{\lambda^{\prime}\bm{k}^{\prime}}-\varepsilon_{\lambda \bm{k}}+\hbar \omega_{\bf{K}})  \delta(\bm{k}^{\prime}-\bm{k}+\bm{q})\\
	\end{aligned}
\end{equation}
\begin{equation}
	\begin{aligned}
		G_{\bf{K}}&=\frac{1}{A}\sum_{\lambda, \lambda^{\prime}}\sum_{\bm{k},\bm{k^{\prime}}} P^{\mathrm{AB},\bf{K}}_{\lambda \bm{k} \lambda^{\prime} \bm{k}^{\prime}} f_{\lambda}(,\bm{k})(1-f_{\lambda^{\prime}}(\bm{k}^{\prime}))\\
		&=\sum_{\lambda, \lambda^{\prime}}\frac{\left\langle \mathrm{D}_{\bf{K}}^2 \right\rangle_F  n_{\bf{K}}}{4 \pi \rho\omega_{\bf{K}} }\int f_{\lambda}(\varepsilon_{\lambda \bm{k}}) N(\varepsilon_{\lambda \bm{k}}) d\varepsilon_{\lambda \bm{k}}\\
		&\quad \times \int d^2\bm{k}^{\prime} \left[ 1 \pm \cos (\theta_{\bm{k},\bm{k}^{\prime}})\right]  (1-f_{\lambda^{\prime}}(\bm{k}^{\prime})) \delta (\varepsilon_{\lambda^{\prime}\bm{k}^{\prime}}-\varepsilon_{\lambda \bm{k}}-\hbar \omega_{\bf{K}})  \delta(\bm{k}^{\prime}-\bm{k}-\bm{q})\\
	\end{aligned}
\end{equation}
\end{subequations}
Using Eqs. (SIII.3)-(SIII.4), the total balance between the optical phonon emission and absorption rate is given by $R^{\mathrm{Net}}_{\eta}=R_{\eta}-G_{\eta}$.

In Eq. (9), $R^{Net}_{M,\eta}=R_{\eta}-G_{\eta}$ denotes the total balance between the optical phonon emission and absorption rate per number of phonon modes.
\begin{subequations}
\begin{equation}
	\begin{aligned}
		R_{M,\bf{\eta}}&=\frac{1}{A}\sum_{\lambda, \lambda^{\prime}}\sum_{\bm{k},\bm{k^{\prime}}} P^{\mathrm{EM},\eta}_{\lambda \bm{k} \lambda^{\prime} \bm{k}^{\prime}} f_{\lambda}(\bm{k})(1-f_{\lambda^{\prime}}(\bm{k}^{\prime}))/M_{\eta}^{-}(\lambda\bm{k})\\
	\end{aligned}
\end{equation}
\begin{equation}
	\begin{aligned}
		G_{M,\bf{\eta}}&=\frac{1}{A}\sum_{\lambda, \lambda^{\prime}}\sum_{\bm{k},\bm{k^{\prime}}} P^{\mathrm{AB},\eta}_{\lambda \bm{k} \lambda^{\prime} \bm{k}^{\prime}} f_{\lambda}(\bm{k})(1-f_{\lambda^{\prime}}(\bm{k}^{\prime}))/M_{\eta}^{+}(\lambda\bm{k})\\
	\end{aligned}
\end{equation}
\end{subequations}
Here, $M_{\eta}^{-}(\lambda\bm{k})$ and $M_{\eta}^{+}(\lambda\bm{k})$ are  the number of $\eta$-phonon modes $(\bm{q})$ per unit area that participate the phonon emission and absorption processes for carries state $(\lambda, \bm{k})$, respectively.
\begin{equation}
	\begin{aligned}
		M_{\eta}^{\pm}(\lambda\bm{k})&=\left. \frac{a_{\eta}}{A} \left( \pi |\bm{q}|_{\mathrm{max}}^2-\pi |\bm{q}|_{\mathrm{min}}^2 \right) \right/ |\Delta \bm{q}| \\
		&=\left. \frac{a_{\eta}}{A} \left| \pi \left(\frac{2\varepsilon_{\lambda \bm{k}} \pm \hbar \omega_{\eta}}{\hbar v_F} \right)^2-\pi \left( \frac{\hbar \omega_{\eta}}{\hbar v_F}\right)^2 \right| \right/ \left\lbrace  \frac{(2 \pi )^2}{A}\right\rbrace \\
		&=  \frac{a_{\eta}}{4 \pi}\left|\left\lbrace\left(\frac{2\varepsilon_{\lambda \bm{k}} \pm \hbar \omega_{\eta}}{\hbar v_F} \right)^2-\left( \frac{ \omega_{\eta}}{ v_F}\right)^2 \right\rbrace\right| 
	\end{aligned}
\end{equation}
In this case, $a_{\bf{\Gamma}}=1$ for $\Gamma$-LO and $\Gamma$-TO phonons, and $a_{\bf{K}}=2$ for $K$ phonon. The factor of $a_{\bf{K}}=2$ represents the degenerate phonon valleys at the $K$ and $K^{\prime}$ points. Using Eqs. (SIII.5), the total balance between the optical phonon emission and absorption rate per number of phonon modes is given by $R^{\mathrm{Net}}_{M,\eta}=R_{M,\eta}-G_{M,\eta}$.
	\newpage

\section{Pump power injected into graphene sample considering saturable absorption} 
In this section, we present the derivation of the pump intensity $\mathcal{F}_{\mathrm{ab}}$ injected into the graphene sample, considering the multiple reflections inside the substrate and the saturable absorption (SA) effect. The SA is an extreme nonlinear phenomenon that consists of the quenching of the optical absorption under high-intensity illumination. Following Marini et al.\cite{Marini2017a}{\tiny }, we introduce the derivation of saturable absorption coefficient $\alpha_{\mathrm{inter}}$ in graphene. Thereafter, we explain the derivation of the absorbed pump intensity $\mathcal{F}_{\mathrm{ab}}$ by graphene on the substrate at an oblique incidence angle using $\alpha_{\mathrm{inter}}$. 

We study the response of a single electron in graphene under an in-plane x-direction applied field $\mathbf{E}(t)=E_0 \mathrm{e}^{-i \omega t}\hat{\mathbf{x}}$. The extended Bloch equations describing the temporal variation in the interband coherence $\rho_{\mathbf{k}}$ and population difference $n_{\mathbf{k}}$ in photoexcited graphene are as follows: 
\begin{subequations}
\begin{equation}
	\dot{\rho}_{\mathbf{k}}(t)=-\frac{i}{2} \dot{\theta}_{\mathbf{k}}(t) n_{\mathbf{k}}(t) \mathrm{e}^{2 i \Omega_{\mathbf{k}}(t)}-\frac{\rho_{\mathbf{k}}(t)}{\tau_{\mathrm{ie}}},
\end{equation}
\begin{equation}
	\dot{n}_{\mathbf{k}}(t)=2 \dot{\theta}_{\mathbf{k}}(t) \operatorname{Im}\left\{\rho_{\mathbf{k}}(t) \mathrm{e}^{-2 i \Omega_{\mathbf{k}}(t)}\right\}-\frac{n_{\mathbf{k}}(t)}{\tau_{\mathrm{ie}}},
\end{equation}
\end{subequations}
where $\pi=\hbar \mathbf{k}=\hbar k(\cos \phi, \sin \phi)$ is the electron momentum, the global dynamical phase  $\Omega_{\mathbf{k}}(t)$ is defined as $\Omega_{\mathbf{k}}(t)=v_{\mathrm{F}} \int|\mathbf{k}+(e / \hbar c) \mathbf{A}(t)| d t$, and $\theta_{\mathbf{k}}(t)=\operatorname{atan}\left\{k_{y} /\left[k_{x}+(e / \hbar c) A(t)\right]\right\}$ is the time-dependent directional angle of the electron quasimomentum $\mathbf{\pi}(t)=\hbar \mathbf{k}+(e / c) \mathbf{A}(t)$. The vector potential $\mathbf{A}(t)$ is $\mathbf{A}(t)=-\int \mathbf{E}(t) dt$. In the near-resonant condition, the optical momentum is negligible [i.e., $\hbar k \gg (e/c)A(t)$] and does not significantly affect the interband dynamics. In this approximation, Eqs. (SIV.1) is reduced to 
\begin{subequations}
\begin{equation}
	\dot{\Gamma}_{\mathbf{k}}=-\left(\frac{1}{\tau_{\mathrm{ie}}}+2 i \omega_{0}\right) \Gamma_{\mathbf{k}}-\frac{i e}{\hbar k} \operatorname{Re}\left\{E_{0} \mathrm{e}^{-i \omega t}\right\} \sin \phi n_{\mathrm{k}}
\end{equation}
\begin{equation}
	\dot{n}_{\mathrm{k}}=-\frac{1}{\tau_{\mathrm{ie}}}\left[n_{\mathbf{k}}-\mathcal{N}\right]+\frac{4 e}{\hbar k} \operatorname{Re}\left\{E_{0} \mathrm{e}^{-i \omega t}\right\} \sin \phi \operatorname{Im}\left\{\Gamma_{\mathbf{k}}\right\}.
\end{equation}
\end{subequations}
In this case, $\Gamma_{\mathbf{k}}(t)=\rho_{\mathbf{k}}(t) \mathrm{e}^{-2 i \omega_{0} t}, \omega_{0}=v_{\mathrm{F}} |\mathbf{k}|$, $\mathcal{N}(k,T_{\mathrm{e}})=\mathcal{F}(\mathbf{k},T_{\mathrm{e}})-\mathcal{F}(-\mathbf{k},T_{\mathrm{e}})$, where $\mathcal{F}(\mathbf{k}.T_{\mathrm{e}})=1/\{1+\exp[(v_F \hbar |\mathbf{k}|-\mu)/k_B T_{\mathrm{e}}]\}$. A phenomenological relaxation time $\tau_{\mathrm{ie}}$ is introduced, which encompasses the effect of numerous ultrafast decay channels for the out-of-equilibrium electrons into hot carriers and phonons. The steady-state ansatz for the Bloch equation is given by 
\begin{subequations}
\begin{equation}
	\Gamma_{\mathbf{k}}(t)=\Gamma_{\mathbf{k}}^{+} \mathrm{e}^{i \omega t}+\Gamma_{\mathbf{k}}^{-} \mathrm{e}^{-i \omega t}
\end{equation}
\begin{equation}
	n_{\mathbf{k}}(t)=n_{\mathbf{k}}^{(0)}+\operatorname{Re}\left\{n_{\mathbf{k}}^{(2)} \mathrm{e}^{-2 i \omega t}\right\}
\end{equation}
\end{subequations}
Using these expressions and neglecting the higher-harmonic terms, Eqs. (SIV.2) leads to 
\begin{subequations}
\begin{equation}
	n_{\mathbf{k}}^{(0)}=\mathcal{N}+4 \xi \operatorname{Im}\left\{\frac{1-i \omega \tau_{\mathrm{ie}}}{1-i \omega_{+} \tau_{\mathrm{ie}}} \Gamma_{\mathbf{k}}^{-}\right\}
\end{equation}
\begin{equation}
	n_{\mathbf{k}}^{(2)}=\frac{-4 i \xi(1-i \omega \tau_{\mathrm{ie}}) \Gamma_{\mathbf{k}}^{-}}{(1-2 i \omega \tau_{\mathrm{ie}})\left(1-i \omega_{+} \tau_{\mathrm{ie}}\right)}
\end{equation}
\begin{equation}
	\Gamma_{\mathbf{k}}^{+}=-\frac{1+i \omega_{-} \tau_{\mathrm{ie}}}{1+i \omega_{+} \tau_{\mathrm{ie}}} \Gamma_{\mathbf{k}}^{-*}
\end{equation}
\begin{equation}
	\Gamma_{\mathbf{k}}^{-}=\frac{(-i \xi / 2)}{1-i \omega_{-} \tau_{\mathrm{ie}}}\left(n_{\mathbf{k}}^{(0)}+\frac{1}{2} n_{\mathbf{k}}^{(2)}\right),
\end{equation}
\end{subequations}
where $\xi=\left(e \tau_{\mathrm{ie}} E_{0} / \hbar k\right) \sin \phi$ and $\omega_{\pm}=\omega \pm 2 \omega_{0}$. The macroscopic interband current density depending on the light intensity $I_0=(c/2 \pi) |E_0|^2$ at the electronic temperature $T_{\mathrm{e}}$ is determined by
\begin{equation}
	\mathbf{J}_{\text {inter }}(t)=-\frac{2 e v_{\mathrm{F}}}{\pi^{2}} \operatorname{Re}\left\{i \mathrm{e}^{-i \omega t} \int \sin \phi\left[\Gamma_{\mathbf{k}}^{-}-\Gamma_{\mathbf{k}}^{+*}\right] d^{2} \mathbf{k}\right\} \hat{\mathbf{x}},
\end{equation}
where $\Gamma_{\mathbf{k}}^{-}-\Gamma_{\mathbf{k}}^{+*}=-\frac{\operatorname{ie\tau_{\mathrm{ie}}} E_{0} \sin \phi(1-i \omega \tau_{\mathrm{ie}})}{2 \hbar k\left(1-i \omega_{+} \tau_{\mathrm{ie}}\right)\left(1-i \omega_{-} \tau_{\mathrm{ie}}\right)}\left[2 n_{\mathbf{k}}^{(0)}+n_{\mathbf{k}}^{(2)}\right]$, and
\begin{subequations}
\begin{equation}
	n_{\mathbf{k}}^{(0)}=\frac{\mathcal{N}}{1+2 \xi^{2} \operatorname{Im}\left\{\frac{i(1-i \omega \tau_{\mathrm{ie}})}{\left(1-i \omega_{+} \tau_{\mathrm{ie}}\right)\left(1-i \omega_{-} \tau_{\mathrm{ie}}\right)}\left[1-\frac{\xi^{2}(1-i \omega \tau_{\mathrm{ie}})}{(1-2 i \omega \tau_{\mathrm{ie}})\left(1-i \omega_{+} \tau_{\mathrm{ie}}\right)\left(1-i \omega_{-} \tau_{\mathrm{ie}}\right)+\xi^{2}(1-i \omega \tau_{\mathrm{ie}})}\right]\right\}}
\end{equation}
\begin{equation}
	n_{\mathbf{k}}^{(2)}=\frac{-2 \xi^{2}(1-i \omega \tau_{\mathrm{ie}}) \mathcal{N} /\left[(1-2 i \omega \tau_{\mathrm{ie}})\left(1-i \omega_{+} \tau_{\mathrm{ie}}\right)\left(1-i \omega_{-} \tau_{\mathrm{ie}}\right)+\xi^{2}(1-i \omega \tau_{\mathrm{ie}})\right]}{1+2 \xi^{2} \operatorname{Im}\left\{\frac{i(1-i \omega \tau_{\mathrm{ie}})}{\left(1-i \omega_{+} \tau_{\mathrm{ie}}\right)\left(1-i \omega_{-} \tau_{\mathrm{ie}}\right)}\left[1-\frac{\xi^{2}(1-i \omega \tau_{\mathrm{ie}})}{(1-2 i \omega \tau_{\mathrm{ie}})\left(1-i \omega_{+} \tau_{\mathrm{ie}}\right)\left(1-i \omega_{-} \tau_{\mathrm{ie}}\right)+\xi^{2}(1-i \omega \tau)}\right]\right\}}.
\end{equation}
\end{subequations}
Subsequently, by expressing the integral over the reciprocal space in polar coordinates, the following is obtained:
\begin{equation}
	\mathbf{J}_{\text {inter }}(t)=-\frac{8 e^{2} v_{\mathrm{F}} \tau_{\mathrm{ie}}}{\pi^{2} \hbar} \operatorname{Re}\left\{E_{0} \mathrm{e}^{-i \omega t}(1-i \omega \tau_{\mathrm{ie}}) \int_{0}^{\pi / 2} d \phi \int_{0}^{\infty} d k \frac{\sin ^{2} \phi\left[2 n_{\mathrm{k}}^{(0)}+n_{\mathrm{k}}^{(2)}\right]}{2\left(1-i \omega_{+} \tau_{\mathrm{ie}}\right)\left(1-i \omega_{-} \tau_{\mathrm{ie}}\right)}\right\} \hat{\mathrm{x}}.
\end{equation}
Using the interband current, the interband absorption coefficient is determined as the ratio of the time-averaged absorbed power over an optical cycle to the incident intensity $I_0$:
\begin{equation}
	\alpha_{\text {inter }} (I_0) \equiv \frac{\int_{-\pi / \omega}^{+\pi / \omega} \mathbf{J}_{\text {inter }}(t) \cdot \mathbf{E}(t) d t}{(2 \pi / \omega) I_0}.
\end{equation}
Although the above results were obtained under the CW illumination conditions, these are also applicable to commonly used optical pulses that have a large duration compared to the optical period.

Taking into account the SA for the interband transition by the pump irradiation, the transmission and reflection coefficients of the s-polarized pump pulse incident on the system of layer i/graphene/layer j from layer i, as illustrated in Fig. S1, are calculated by 
\begin{subequations}
\begin{equation}
	\begin{aligned}
		t_{\mathrm{ij}}^s(I_0, \gamma_{\mathrm{ij}})=\frac{2 \epsilon_{i}^{1 / 2} (\omega_{\mathrm{pump}}) \cos \theta_{\mathrm{i}}}{\alpha_{\text {inter }} (\gamma_{\mathrm{ij}} I_0)+(\epsilon_{\mathrm{j}}(\omega_{\mathrm{pump}})-\epsilon_{\mathrm{i}}(\omega_{\mathrm{pump}}) \sin^2 \theta_{\mathrm{i}})^{1 / 2}+\epsilon_{\mathrm{i}} (\omega_{\mathrm{pump}})^{1 / 2} \cos \theta_{\mathrm{i}}},
	\end{aligned}
\end{equation}
\begin{equation}
	\begin{aligned}
		r_{\mathrm{ij}}^s(I_0, \gamma_{\mathrm{ij}})=-\frac{\alpha_{\text {inter }} (\omega,\gamma_{\mathrm{ij}} I_0,T_{\mathrm{e}})+(\epsilon_{\mathrm{j}}(\omega_{\mathrm{p}})-\epsilon_{\mathrm{i}}(\omega_{\mathrm{p}}) \sin^2 \theta_{\mathrm{i}})^{1 / 2}-\epsilon_{\mathrm{i}}^{1 / 2}(\omega_{\mathrm{p}}) \cos \theta_{\mathrm{i}}}{\alpha_{\text {inter }} (\omega,\gamma_{\mathrm{ij}} I_0,T_{\mathrm{e}})+(\epsilon_{\mathrm{j}}(\omega_{\mathrm{p}})-\epsilon_{\mathrm{i}}(\omega_{\mathrm{p}}) \sin^2 \theta_{\mathrm{i}})^{1 / 2}+\epsilon_{\mathrm{i}}^{1 / 2}(\omega_{\mathrm{p}}) \cos \theta_{\mathrm{i}}}.
	\end{aligned}
\end{equation}
\end{subequations}
In this case, the pump pulse irradiates the graphene from layer i with the incidence angle of $\theta_{\mathrm{i}}$ and transmits it to layer j with the angle $\theta_{\mathrm{j}}$. Moreover, $\gamma_{\mathrm{ij}}$ is the correction factor. Although $\alpha_{\text {inter }} (\omega,I_0,T_{\mathrm{e}})$ is appropriate for the case in which the optical pump pulse excites the suspended graphene at the normal incidence angle, the saturation behavior will change when graphene on a substrate is excited by a pump pulse at an oblique incidence angle, where the injected pump power becomes smaller by a factor of $\gamma_{\mathrm{ij}}$.
The corresponding transmittance and reflectance are determined by 
\begin{subequations}
\begin{equation}
	\begin{aligned}
		T_{\mathrm{ij}}^s (I_0, \gamma_{\mathrm{ij}})=|t_{\mathrm{ij}}^s(I_0, \gamma_{\mathrm{ij}})|^2 \frac{\epsilon_{\mathrm{j}}^{1 / 2}(\omega_{\mathrm{p}}) \cos \theta_{\mathrm{j}}}{\epsilon_{\mathrm{i}}^{1 / 2}(\omega_{\mathrm{p}}) \cos \theta_{\mathrm{i}}},
	\end{aligned}
\end{equation}
\begin{equation}
	\begin{aligned}
		R_{\mathrm{ij}}^s(I_0, \gamma_{\mathrm{ij}})=|r^s_{\mathrm{ij}}(I_0, \gamma_{\mathrm{ij}})|^2.
	\end{aligned}
\end{equation}
\end{subequations}
Using Eq. (SIV.10), the absorption of the pump pulse by the graphene layer is provided by
\begin{equation}
	\begin{aligned}
		A^{s}_{\mathrm{ij}}(I_0, \gamma_{\mathrm{ij}})=1-T^s_{\mathrm{ij}}(I_0, \gamma_{\mathrm{ij}})-R^s_{\mathrm{ij}} (I_0, \gamma_{\mathrm{ij}}).
	\end{aligned}
\end{equation}
The correction factor $\gamma_{\mathrm{ij}}$ is calculated by the ratio of the absorption coefficient
\begin{equation}
	\begin{aligned}
		\gamma_{\mathrm{ij}}=\frac{A^{s}_{\mathrm{ij}}(I_0, \gamma_{\mathrm{ij}})}{\alpha_{\text {inter }} (I_0)},
	\end{aligned}
\end{equation}
and can be determined self-consistently. Using the converged $\gamma^*_{\mathrm{ij}}$, the transmittance, reflectance, and absorption coefficients in the experimental condition are obtained by
\begin{subequations}
\begin{equation}
	\begin{aligned}
		T^{s*}_{\mathrm{ij}}(I_0)=T^{s}_{\mathrm{ij}}(I_0, \gamma^*_{\mathrm{ij}}),
	\end{aligned}
\end{equation}
\begin{equation}
	\begin{aligned}
		R^{s*}_{\mathrm{ij}}(I_0)=R^{s}_{\mathrm{ij}}(I_0, \gamma^*_{\mathrm{ij}}),
	\end{aligned}
\end{equation}
\begin{equation}
	\begin{aligned}
		A^{s*}_{\mathrm{ij}}(I_0)=A^{s}_{\mathrm{ij}}(I_0, \gamma^*_{\mathrm{ij}}).
	\end{aligned}
\end{equation}
\end{subequations}
The envelope function of the pump pulse considering the $n^{\mathrm{th}}$ multiple reflections inside the substrate is given by
\begin{equation}
	\begin{aligned}
		\mathcal{I}(t)=\sum_{n=0} \mathcal{I}_n(t+n\Delta T)
	\end{aligned}
\end{equation}
In this case, $\mathcal{I}_0(t)$ represents the incident pump pulse, which is assumed to have hyperbolic secant form $\mathcal{I}_0(t)=\left(F_{\mathrm{0}} / 2 \tau_{\mathrm{pump}}\right) \operatorname{sech}^{2}\left(t / \tau_{\mathrm{pump}}\right)$, where $F_{\mathrm{0}}$ is the fluence and $2\tau_{\mathrm{pump}}$ is the pulse duration. $\mathcal{I}_n(t)=\left(F_{\mathrm{n}} / 2 \tau_{\mathrm{pump}}\right) \operatorname{sech}^{2}\left(t / \tau_{\mathrm{punmp}}\right)$ represents the $n^{th}$ reflection of the incident pump pulse and $F_{\mathrm{n}}$ is the fluence of the $n^{th}$ reflection pulse. $\Delta T$ is the time delay owing to one round trip in the substrate. Using Eq. (SIV.13) and $I_0=F_{\mathrm{0}}/ 2 \tau_{\mathrm{pump}}$, $F_n$ for $n \geq 1$ is obtained by 
\begin{subequations}
	\begin{equation}
		F_1=F_0 T^{s*}_{\mathrm{12}}(F_{\mathrm{0}} / 2 \tau_{\mathrm{pump}})R^{s}_{23},
	\end{equation}
	\begin{equation}
		F_n=F_{n-1} R^{s*}_{21}(F_{\mathrm{n-1}} / 2 \tau_{\mathrm{pump}})R^{s}_{23},\; (\text{for}\; n \geq 2). 
	\end{equation}
\end{subequations}
In the above, $R^{s}_{23}$ is the reflectance of the pump pulse incident at the substrate (layer 2) /$N_2$ purged (layer 3) interface from the substrate ($\alpha_{\text {inter}} (I_0)=0$ in Eq. (SIV.13b)). Using Eqs. (SIV.13), (SIV.14), and (SIV.15), the absorbed pump intensity $\mathcal{F}_{ab}(t)$ is determined by
\begin{equation}
	\begin{aligned}
		\mathcal{I}_{ab}(t)=\mathcal{I}_0(t) A^{s*}_{\mathrm{12}}(F_{\mathrm{0}} / 2 \tau_{\mathrm{pump}})+\sum_n \mathcal{I}_n(t+ n \Delta t) A^{s*}_{\mathrm{21}}(F_{\mathrm{n}} / 2 \tau_{\mathrm{pump}}).
	\end{aligned}
\end{equation}

Figures S2(a)--(f) depict the pump intensity dependence of $\alpha_{\mathrm{inter}}$ and $A^{s}_{\mathrm{ij}}$ for various $\tau_{\mathrm{ie}}$ and $T_e$ calculated using Eqs. (SIV.8) and (SIV.13c).
Figure S3 (a) and (b) present the saturated pump intensities $I_s$ for $\alpha_{\mathrm{inter}}$ and $A^{s*}_{\mathrm{12}}$, respectively, where $I_s$ is defined as
 $\alpha_{\mathrm{inter}}(I_s)=(1/2) \alpha_{\mathrm{inter}}(0) $ and  $A^{s*}_{\mathrm{12}}(I_s)=(1/2) A^{s*}_{\mathrm{12}}(0) $. Figure S4 shows the absorbed pump fluence in graphene with $|\varepsilon_{\mathrm{F}}|=$0.15 and $0.43\,\mathrm{eV}$, calculated using Eq. (SIV.16). 
\begin{figure}[b]
	\centering
	\includegraphics[width=9cm, bb=0 0 311 401]{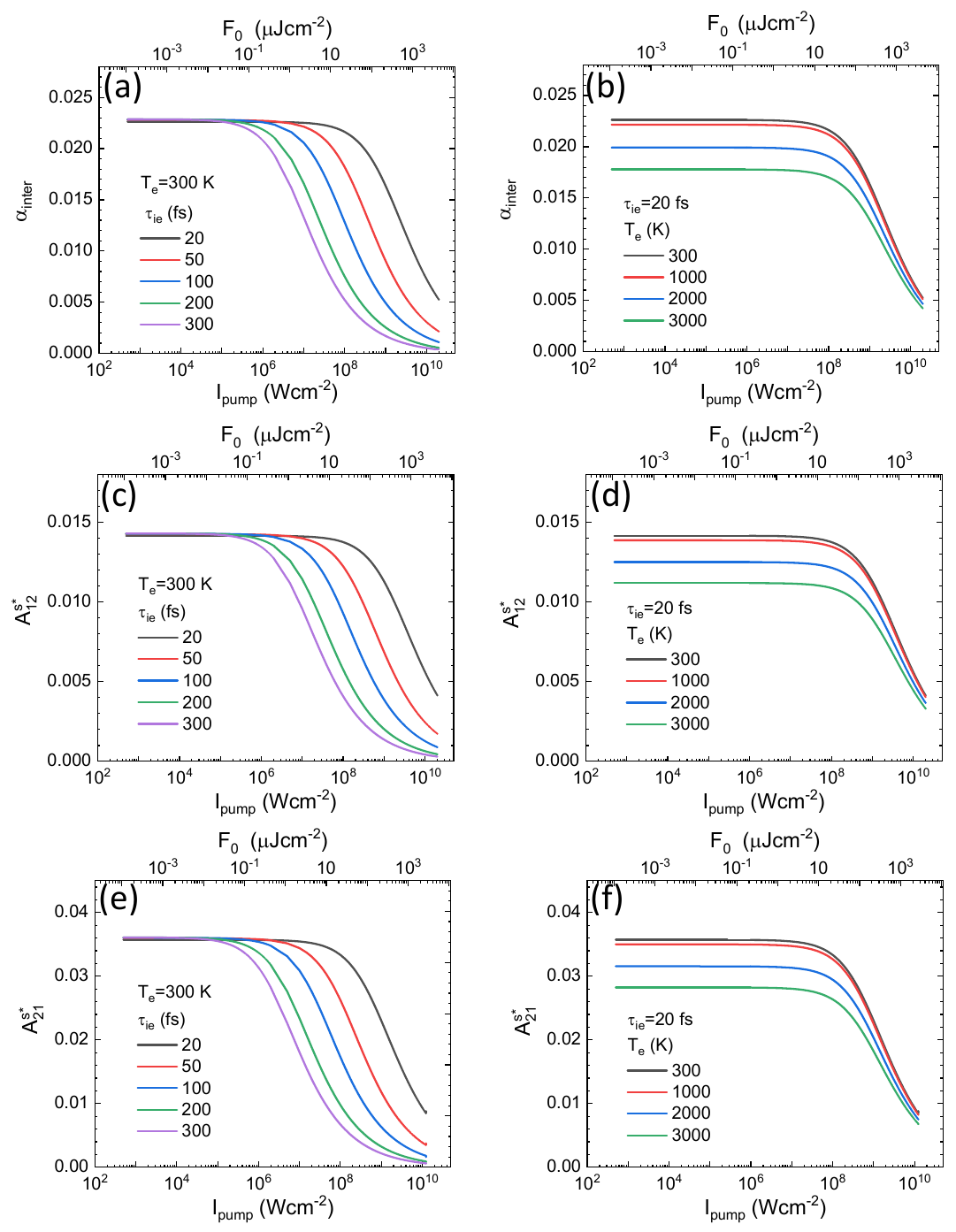}
	\caption{\label{figS2} (a)--(f) Pump intensity dependence $I_{\mathrm{pump}}$ of $\alpha_{\mathrm{inter}}$ and $A^{s}_{\mathrm{ij}}$ at $\theta=60^{\circ}$ in heavily doped graphene with $|\varepsilon_{\mathrm{F}}|=0.43\,\mathrm{eV}$, assuming $\epsilon_2=2.4$ for various $\tau_{\mathrm{ie}}$ and $T_e$. The $F_0=I_0 \times 2\tau_{\mathrm{pump}}$ on the upper horizontal axis was calculated by assuming $2\tau_{\mathrm{pump}}=$220\,fs in the OPTP experiment.}
\end{figure}
\begin{figure}
	\centering
	\includegraphics[width=9cm, bb=0 0 312 147]{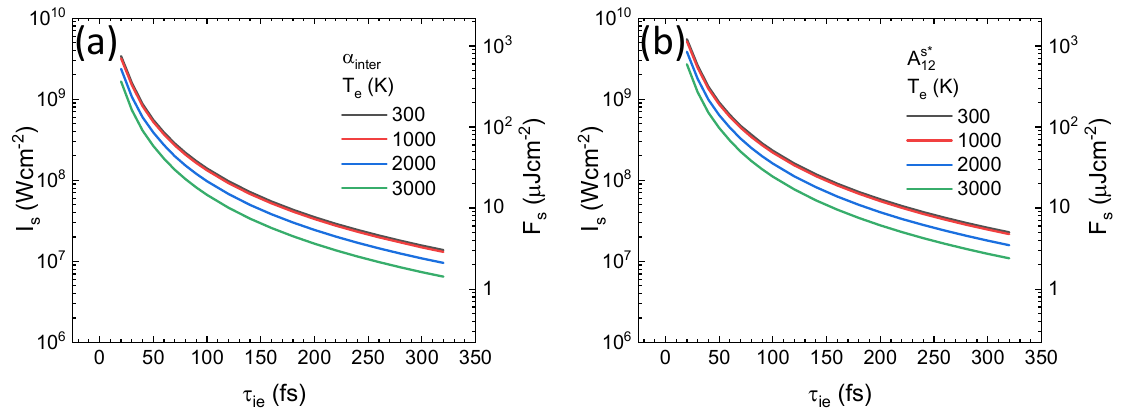}
	\caption{\label{figS3} $T_e$ dependence of saturated pump intensity $I_s$ of heavily doped graphene with $|\varepsilon_{\mathrm{F}}|=0.43\,\mathrm{eV}$, assuming $\epsilon_2=2.4$ for (a) $\alpha_{\mathrm{inter}}$ and (b) $A^{s}_{\mathrm{ij}}$ at $\theta=60^{\circ}$. The $F_{\mathrm{s}}=I_{\mathrm{s}} \times 2\tau_{\mathrm{pump}}$ on the right vertical axis was calculated by assuming $2\tau_{\mathrm{pump}}=$220\,fs in the OPTP experiment}
\end{figure}
\begin{figure}
	\centering
	\includegraphics[width=15cm, bb=0 0 343 91]{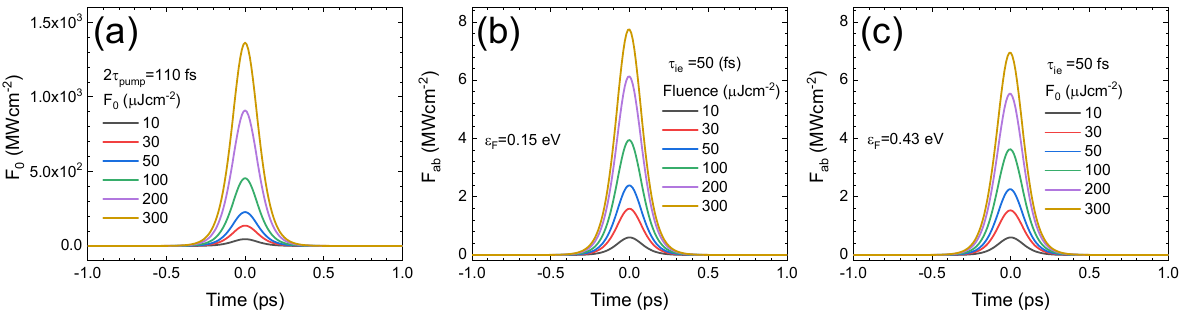}
	\caption{\label{figS4} (a) Envelope function of pump intensity $F_{\mathrm{0}}$  incident on graphene at $\theta=60^{\circ}$. Absorbed pump intensity $F_{\mathrm{ab}}$ in graphene with $|\varepsilon_{\mathrm{F}}|=$(b)0.15 and (c)0.43\,eV at $T_e$=295\,K using $\epsilon_2=2.4$.}
\end{figure}

\clearpage
\section{Calculation of the transient THz reflection change from optical conductivity}
In this section, we explain the calculation procedure of the transient reflection change $\Delta E_{\mathrm{r}}(\tau_1)/E_0$ from $\sigma(\omega, \tau_1)$, calculated using the iterative solution of the BTE and the four-temperature model. 
The reflected THz electric field in the time domain, $E_{\mathrm{r}}^{\mathrm{s}}\left(\tau_{2}, \tau_{1}\right)$, where $\tau_{2}$ is the probe trigger delay, is determined by the inverse Fourier transformation of the reflected THz electric field in the frequency domain, $E_{\mathrm{r}}^{\mathrm{s}}\left(\omega_{\mathrm{THz}}, \tau_{1}\right)$:
\begin{equation}
	\begin{aligned}
		E_{\mathrm{r}}^{\mathrm{s}}\left(\tau_{2}, \tau_{1}\right) &=\int E_{\mathrm{r}}^{\mathrm{s}}\left(\omega_{\mathrm{THz}}, \tau_{1}\right) e^{i \omega_{\mathrm{THz}} \tau_{2}} d \omega_{\mathrm{THz}}\\
		&=\int E_{\mathrm{i}}^{\mathrm{s}}\left(\omega_{\mathrm{THz}}\right) r_{\mathrm{s}}\left(\omega_{\mathrm{THz}}, \tau_{1}\right) e^{i \omega_{\mathrm{THz}} \tau_{2}} d \omega_{\mathrm{THz}},
	\end{aligned}
\end{equation}
where $E_{\mathrm{i}}^{\mathrm{s}}\left(\omega_{\mathrm{THz}}\right)$ is the electric field of the incident THz pulse in the frequency domain and $r_s^{\prime}\left(\omega_{\mathrm{THz}}, \tau_{1}\right)$ is the refection coefficient of the THz probe by the photoexcited graphene at $\tau_1$, which is calculated as a function of $\sigma(\omega, \tau_1)$ by Eq. (SII.1).
The normalized reflection change $\Delta E_{\mathrm{r}}^{\mathrm{s}}\left(\tau_{2}, \tau_{1}\right)/E_{\mathrm{r}}^{\mathrm{s}}\left(\tau_{2}\right)$ $\Delta E_{\mathrm{r}}(\tau_1)/E_0$ as a function of the probe trigger delay $\tau_2$ at $\tau_1$ is expressed by
\begin{equation}
	\frac{\Delta E_{\mathrm{r}}(\tau_1)}{E_0} \equiv \frac{\Delta E_{\mathrm{r}}^{\mathrm{s}}\left(\tau_{2}, \tau_{1}\right)}{E_{\mathrm{r}}^{\mathrm{s}}\left(\tau_{2}\right)}=\frac{E_{\mathrm{r}}^{\mathrm{s}}\left(\tau_{2}, \tau_{1}\right)-E_{\mathrm{r}}^{\mathrm{s}}\left(\tau_{2}\right)}{E_{\mathrm{r}}^{\mathrm{s}}\left(\tau_{2}\right)}.
\end{equation}
In the above, $E_{\mathrm{r}}^s\left(\tau_{2}\right)$ is the reflected THz field through the graphene sample without pump fluence. 
We define the transient reflectivity $\Delta E_{\mathrm{r}}(\tau_1)/E_0 \equiv \Delta E_{\mathrm{r}}^{\mathrm{s}}(\tau_{2}, \tau_{1})/E_{\mathrm{r}}^{\mathrm{s}}(\tau_{2})$ at $\tau_2=$0\,ps when the peak amplitude of $E_{\mathrm{r}}^s\left(\tau_{2}\right)$ takes the maximum amplitude.

Figures S5(a) and (b) depict the temporal evolution of $T_e$ and $T_{\eta}$ of the photoexcited graphene, and the temporal waveforms and Fourier spectra of the THz probe pulse, respectively, used in the calculation of $\sigma (\omega,\tau_1)$\cite{Yamashita2021} in Figs. S5(d)--(f) for $\left\langle D_{\textbf{K}}^{2}\right\rangle_{\mathrm{F}}=193\,\mathrm{eV}$.
The $\sigma (\omega,\tau_1)$ values are plotted only in the frequency range corresponding to the bandwidth of the THz probe, because the numerical error occurs outside the frequency of the bandwidth. $\sigma (\omega,\tau_1)$ is strongly dependent on the waveform of the THz probe pulse, and non-Drude frequency dependence clearly appears at $\tau_1=0.1\,\mathrm{ps}$ when the carrier distribution and scattering rate change very rapidly during the THz probing time owing to the photoexcitation. Figure S6 depict the $\left\langle D_{\textbf{K}}^{2}\right\rangle_{\mathrm{F}}$ dependence of $\Delta E_{\mathrm{r}}(\tau_1)/E_0$ for different $2\tau_{\mathrm{prob}}$ values calculated using Eq. (SIV.2). The $\Delta E_{\mathrm{r}}(\tau_1)/E_0$ reflects the change of the $\sigma (\omega,\tau_1)$ around the center frequency of THz probe pulse and the peak value becomes higher depending on the $\left\langle D_{\textbf{K}}^{2}\right\rangle_{\mathrm{F}}$.
\begin{figure}[h]
	\centering
	\includegraphics[width=10cm, bb=0	0 334 293]{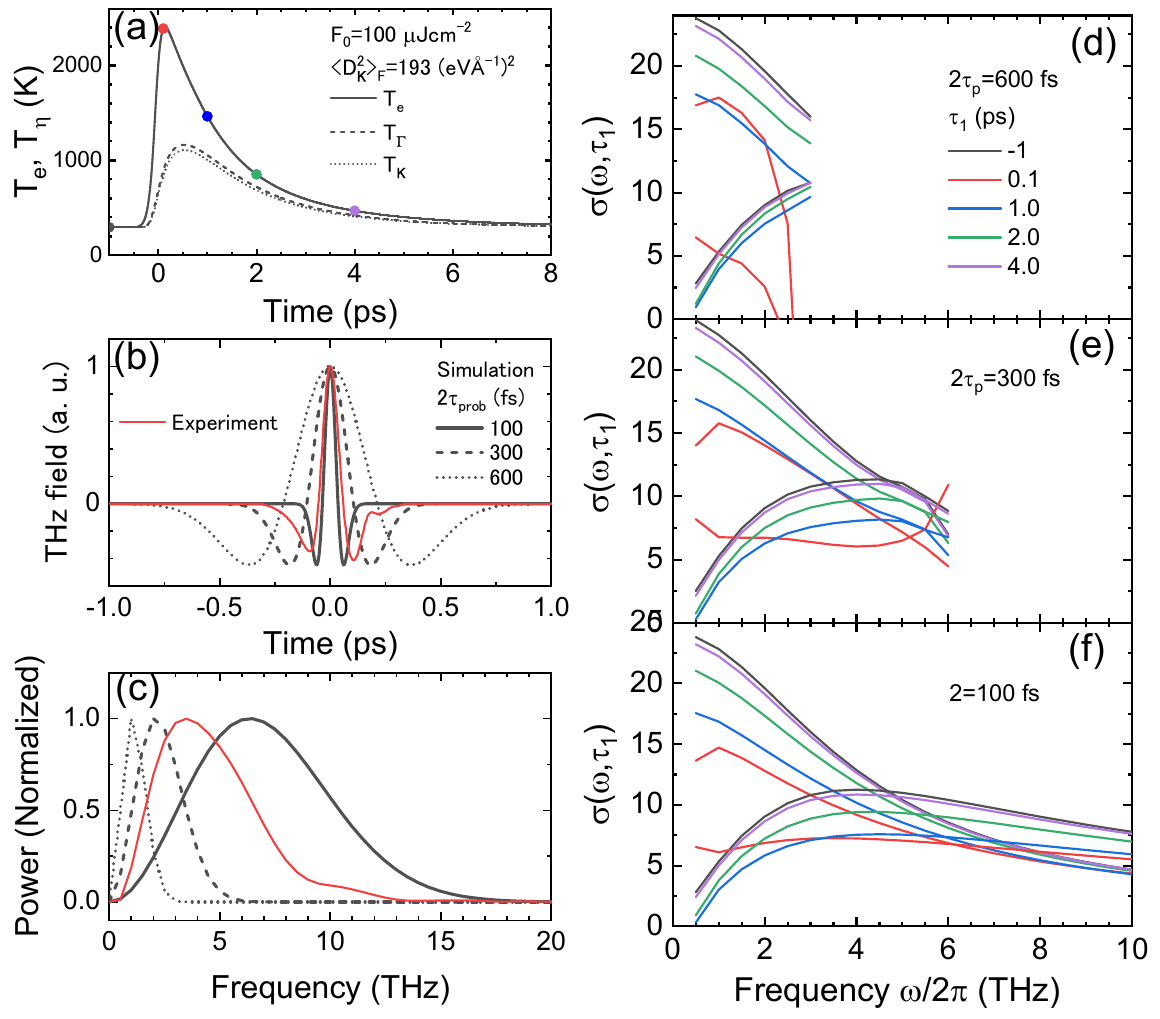}
	\caption{\label{figS5} (a) Temporal evolution of $T_e$ and $T_{\eta}$ of heavily doped graphene with $|\varepsilon_{\mathrm{F}}|=0.43\,\mathrm{eV}$. (b)Temporal waveform of THz probe pulse with $2 \tau_{\mathrm{prob}}=100, 300$ and $600\,\mathrm{(fs)}$ used in simulation. (c) Corresponding Fourier spectrum of THz probe pulse. Temporal evolution of $\sigma (\omega, \tau_1)$ at $\tau_1=-1.0, 0.1, 1.0, 2.0,$ and $4.0\,\mathrm{ps}$ calculated using THz probe with 
	$2 \tau_1$= (d) 600, (e) 300, and (f) 100 fs.}
\end{figure}
\begin{figure}
	\centering
	\includegraphics[width=6.5cm, bb=0 0 217 368]{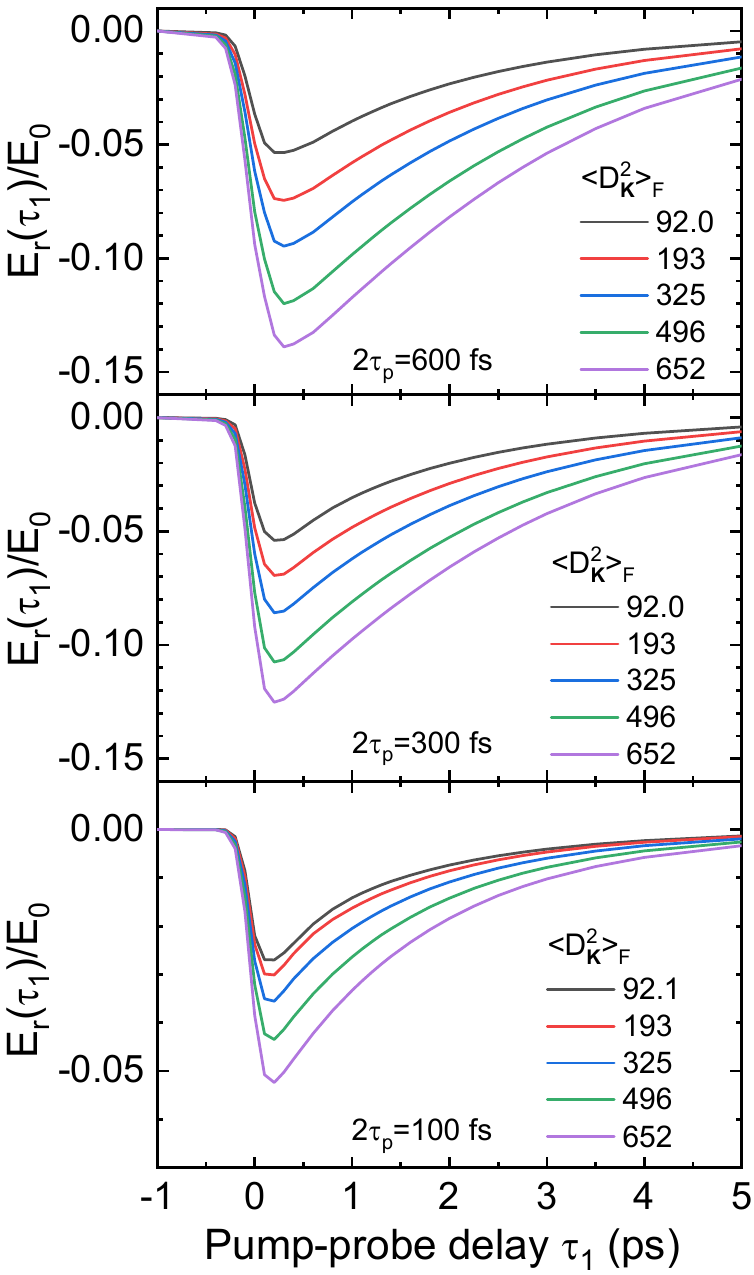}
	\caption{\label{figS6} $\left\langle D_{\textbf{K}}^{2}\right\rangle_{\mathrm{F}}$ dependence of $\Delta E_{\mathrm{r}}(\tau_1)/E_0$} of heavily doped graphene calculated using THz probe pulse with $2\tau_{\mathrm{prob}}=600, 300,$ and $100\,\mathrm{fs}$ for $F_0=100 \mathrm{\mu Jcm^{-2}}$.
\end{figure}

	\clearpage

\bibliographystyle{apsrev4-1.bst}
\bibliography{library}